\documentclass[12pt]{article}
\usepackage{amsfonts,amsmath}
\usepackage{amssymb}
\usepackage[usenames]{color}
\usepackage{mathrsfs}
\usepackage{cancel}
\usepackage{tikz}
\usepackage{bbm}
\usepackage[T2A]{fontenc}
\usepackage{graphicx}
\usepackage[utf8]{inputenc}

\newcommand{\dr}{{{\rm d}}}
\renewcommand{\theequation}{\thesection.\arabic{equation}}

\makeatletter \@addtoreset{equation}{section} \makeatother

{\vspace{3mm} }

\def\*{\star}

\def\E2{\mathbf{E}}

\newcommand{\be}{\begin{equation}}
\newcommand{\ee}{\end{equation}}
\newcommand{\bee}{\begin{eqnarray}}
\newcommand{\beee}{\begin{array}}
\newcommand{\eee}{\end{eqnarray}}
\newcommand{\eeee}{\end{array}}
\newcommand{\gga}{\gamma}

\newcommand{\go}{\omega}

\newcommand{\half}{\frac{1}{2}}

\newcommand{\Ri}{{}\, \widehat{\ast}_R\,}
\newcommand{\Le}{{}\, \widehat{\ast}_L\,}
\newcommand{\hmt}{\vartriangle}

\begin{document}

\begin{flushright}
FIAN/TD/05-25\\
\end{flushright}

\vspace{0.5cm}
\begin{center}
{\large\bf On consistency of the interacting (anti)holomorphic higher-spin sector}

\vspace{1 cm}

\textbf{A.V.~Korybut}\\

\vspace{1 cm}

\textbf{}\textbf{}\\
 \vspace{0.5cm}
 \textit{I.E. Tamm Department of Theoretical Physics,
Lebedev Physical Institute,}\\
 \textit{ Leninsky prospect 53, 119991, Moscow, Russia }\\

\par\end{center}

\begin{center}
\vspace{0.6cm}

\par\end{center}

\vspace{0.4cm}

\begin{abstract}
\noindent In  the recently proposed generating systems for the (anti)holomorphic sector of the 4d higher spin theory \cite{Didenko:2022qga}  and for the off-shell higher spin theory in generic dimension \cite{Didenko:2023vna} locality was achieved due to a peculiar {\it limiting star product}. Even though the generating systems exhibit all-order locality, the product itself encounters uncertainties when functions from specific classes are multiplied. This fact leads to the absence of the Leibniz rule for the differential operator acting on the auxiliary variables $z$ and, hence, its ambiguous definition in the generating equations. We identify the gap in the original proof of consistency associated with this freedom. Nonetheless the generating systems proposed in \cite{Didenko:2022qga}, \cite{Didenko:2023vna} are perfectly consistent as shown by direct computations on the resulting vertices. Considering specific orderings of fields we show that consistency rests on the star-exchange-like identities for the limiting star product formulated and proved here. Connection with the 4d Vasiliev theory is discussed.
\end{abstract}
\newpage

\tableofcontents

\section{Introduction}
Higher spin (HS) gauge theory is a theory that describes interactions of fields of all spins. Such theories are argued to be responsible for dynamics at ultra-high energies \cite{Vasiliev:2016xui}; moreover, they find their application in holography \cite{Klebanov:2002ja} and cosmology \cite{Barvinsky:2015wvz}. Even though there are still unresolved problems in HS theory, there are a series of early observations that point to the fact that HS theory is not a local field theory in the conventional sense. The number of derivatives in its interaction vertices grows with spin, and, since spin is not bounded from above, the number of derivatives grows indefinitely \cite{Bengtsson:1983pd}, \cite{Berends:1984wp}, \cite{Fradkin:1987ks}, \cite{Fradkin:1991iy}. In that perspective, a different notion of locality was developed, promoted as {\it spin-locality}\footnote{The definition of spin-locality unfortunately might change from paper to paper within the literature due to its perturbative nature. A precise connection between the number of spacetime derivatives in vertices in a metric-like formulation and the form of the vertices in the unfolded formalism was established in \cite{Vasiliev:2023yzx}. Explicit expressions for vertices in the metric-like formulation mapped from the unfolded ones are available in \cite{Misuna:2017bjb}, \cite{Tatarenko:2024csa}.}. Spin-locality implies that for a given set of spins, the corresponding vertices have a finite number of derivatives (at least in the lowest orders of perturbation theory). Even though one can come up with a different notion of locality (see \cite{Gelfond:2023fwe}, where a milder notion of locality is introduced that democratically treats holomorphic and antiholomorphic contractions), it is still vital for the theory. It constrains the class of possible field redefinitions, and hence preserves the predictability of the theory because otherwise, by virtue of completely unconstrained non-local field redefinitions, one might make almost any interaction terms vanish.

There are many approaches to HS dynamics available in the literature (see, for example, reviews \cite{Bekaert:2004qos}, \cite{Didenko:2014dwa}, \cite{Ponomarev:2022vjb}). It is worth mentioning that there are approaches to HS dynamics that sacrifice locality already at the cubic level \cite{Aharony:2020omh}. The guiding principle here is the so-called bi-local approach, which guarantees the proper form of the boundary correlators by construction \cite{Das:2003vw} (see also \cite{Jevicki:2015sla}, where the same bi-local approach was studied in the context of boundary thermal CFT). So far, there is only one approach \cite{Vasiliev:1990en} (see also \cite{Vasiliev:1992av}) that covers the full HS dynamics in 4d and does not sacrifice locality from the very beginning. It has the form of a generating system that allows for obtaining consistent nonlinear corrections by solving equations on additional auxiliary variables. The resulting dynamics of HS fields is formulated in the so-called unfolded form \cite{Vasiliev:1988xc}, \cite{Vasiliev:1988sa} (see also \cite{Boulanger:2008up}, \cite{Boulanger:2008kw}, \cite{Misuna:2024dlx}, \cite{Misuna:2024ccj}, \cite{Misuna:2023lpd}, \cite{Misuna:2022cma}, \cite{Misuna:2022zjr}, \cite{Misuna:2019ijn}, \cite{Khabarov:2021djm}, \cite{Iazeolla:2025btr} for recent developments of the formalism and its applications)
\begin{multline}\label{sch1form}
\dr_x \go+\go\ast \go=\Upsilon^\eta(\go,C,C)+\Upsilon^{\bar{\eta}}(\go,C,C)+\\
+\Upsilon^{\eta\eta}(\go,\go,C,C)+\Upsilon^{\eta\bar{\eta}}(\go,\go,C,C)+\Upsilon^{\bar{\eta}\bar{\eta}}(\go,\go,C,C)+\ldots ,
\end{multline} 
\begin{multline}\label{sch0form}
\dr_x C+[\go,C]_\ast=\Upsilon^\eta(\go,C,C)+\Upsilon^{\bar{\eta}}(\go,C,C)+\\+\Upsilon^{\eta\eta}(\go,C,C,C)+\Upsilon^{\eta\bar{\eta}}(\go,C,C,C)+\Upsilon^{\bar{\eta}\bar{\eta}}(\go,C,C,C)+\ldots
\end{multline}
In the unfolded approach the Fronsdal double-traceless fields $\varphi_{\nu_1\ldots \nu_s}$  and their derivatives are packed in the components of the fields $\omega$ and $C$
\begin{equation}
\go(y,\bar{y}|x)=\sum_{m,n}\dr x^\mu \omega_{\mu\, \alpha_1 \ldots \alpha_m,\dot{\alpha}_1 \ldots \dot{\alpha}_n} y^{\alpha_1}\ldots y^{\alpha_m}\bar{y}^{\dot{\alpha}_1}\ldots \bar{y}^{\dot{\alpha}_n}\;\;\; m+n=2(s-1)\,,
\end{equation}
\begin{equation}
C(y,\bar{y}|x)=\sum_{m,n}C_{\alpha_1 \ldots \alpha_m,\dot{\alpha}_1\ldots \dot{\alpha}_n} y^{\alpha_1}\ldots y^{\alpha_m}\bar{y}^{\dot{\alpha}_1}\ldots \bar{y}^{\dot{\alpha}_n}\;\;\; |m-n|=2s\,.
\end{equation}
Both functions $\omega$ and $C$ are endowed with the associative Moyal-product
\begin{equation}\label{Moyal}
f(y,\bar{y})\ast g(y,\bar{y})=f(y,\bar{y})\exp\left\{i\epsilon^{\alpha \beta} \frac{\overleftarrow{\partial}}{\partial y^\alpha}\frac{\overrightarrow{\partial}}{\partial y^\beta}+i\bar{\epsilon}^{\dot{\alpha}\dot{\beta}}\frac{\overleftarrow{\partial}}{\partial \bar{y}^{\dot{\alpha}}}\frac{\overrightarrow{\partial}}{\partial \bar{y}^{\dot{\beta}}}\right\}g(y,\bar{y})\,.
\end{equation}
The linearized version of the equations \eqref{sch1form}, \eqref{sch0form} on the  $AdS_4$ background produces the same dynamics as the Fronsdal equations \cite{Fronsdal:1978rb} which is the content of the famous {\it central on-mass-shell theorem} \cite{Vasiliev:1988sa}, \cite{Vasiliev:1999ba}.  The generating equations of the Vasiliev system schematically have the form
\begin{equation}\label{generate}
\dr_z \Phi^a_n=F^a(\Phi^b_{n-1},\Phi^c_{n-2},\Phi^d_{n-3},\ldots)\,,
\end{equation}
Where $\dr_z$ is the de Rham differential in the auxiliary direction $z$, $\Phi^a_n$ are the so-called master fields for which the upper index specifies the type of the master-field in the Vasiliev equations and the lower index specifies the order of perturbation theory, i.e. each $\Phi^a_n(\underbrace{C,\ldots,C}_n)$ depends on $n$ copies of the $C$-field. A particular solution to the equation \eqref{generate} can be obtained, for example, by virtue of the Poincaré lemma, however one can always adjust it by the solution to homogeneous equation. An unfortunate choice of a particular solution may lead to nonlocal vertices even in the lowest orders of perturbation theory \cite{Skvortsov:2015lja}. 
Nonetheless, techniques for obtaining spin-local vertices from Vasiliev generating system were developed and the corresponding vertices were computed \cite{Vasiliev:2023yzx}, \cite{Gelfond:2023fwe}, \cite{Gelfond:2018vmi},  \cite{Didenko:2018fgx}, \cite{Didenko:2019xzz},  \cite{Gelfond:2019tac}, \cite{Didenko:2020bxd},\cite{Gelfond:2021two}, \cite{Tarusov:2022qpo}. The developed techniques appear to be very universal and were successfully applied to similar problems for HS theory in 3d \cite{Prokushkin:1998bq}, and Coxeter extension of the 4d HS theory \cite{Vasiliev:2018zer} in \cite{Korybut:2022kdx} and \cite{Tarusov:2025qfo} respectively. 

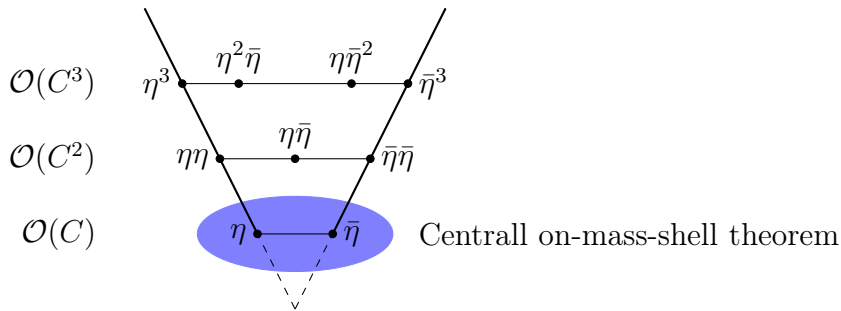
\begin{figure}[h]
\begin{center}
\begin{tikzpicture}
\filldraw[blue!50] (5,3) ellipse (1.3 and 0.5);
\draw[dashed] (5,2) -- (3,6);
\draw[dashed] (5,2) -- (7,6);
\draw[thick] (4.5,3) -- (3,6);
\draw[thick] (5.5,3) -- (7,6);
\filldraw[black] (4.5,3) circle (0.05);
\filldraw[black] (5.5,3) circle (0.05);
\filldraw[black] (4,4) circle (0.05);
\filldraw[black] (6,4) circle (0.05);
\filldraw[black] (5,4) circle (0.05);
\filldraw[black] (3.5,5) circle (0.05);
\filldraw[black] (6.5,5) circle (0.05);
\filldraw[black] (3.5+0.75,5) circle (0.05);
\filldraw[black] (6.5-0.75,5) circle (0.05);
\draw (4.5,3) -- (5.5,3);
\draw (3.5,5) -- (6.5,5);
\draw (4,4) -- (6,4);
\node[left] at (4.5,3) {$\eta$};
\node[right] at (5.5,3) {$\bar{\eta}$};
\node[left] at (4,4) {$\eta\eta$};
\node[right] at (6,4) {$\bar{\eta}\bar{\eta}$};
\node[above] at (5,4) {$\eta\bar{\eta}$};
\node[left] at (3.5,5) {$\eta^3$};
\node[right] at (6.5,5) {$\bar{\eta}^3$};
\node[above] at (3.5+0.75,5) {$\eta^2 \bar{\eta}$};
\node[above] at (6.5-0.75,5) {$\eta \bar{\eta}^2$};
\node[left] at (2.5,5) {$\mathcal{O}(C^3)$};
\node[left] at (2.5,4) {$\mathcal{O}(C^2)$};
\node[left] at (2.5,3) {$\mathcal{O}(C)$};
\node[right] at (6.5,3) {Centrall on-mass-shell theorem};
\end{tikzpicture}
\end{center}
\caption{Vertices in one-form sector of 4d HS theory.} \label{fig:vert}
\end{figure}

Interaction vertices produced by the 4d Vasiliev generating system can be classified as shown in the figure \ref{fig:vert} (see \eqref{sch1form}), where each level corresponds to order of the perturbation theory and $\eta$ and $\bar{\eta}$ are the only coupling constants in the Vasiliev theory. Edges of the cone correspond to the so-called (anti)holomorphic vertices. One can consider a consistent truncation of the full system, leaving only these vertices non-trivial \cite{Vasiliev:1992av}, the so-called {\it self-dual} HS theories. Such HS theories are sometimes called {\it chiral} in literature. Within the framework of the Vasiliev generating system the vertices were obtained in the spin-local form  up to the fifth order on the action level \cite{Gelfond:2021two}. 

Eventually, the highly anticipated breakthrough was made and the generating system for the (anti)holomorphic vertices in self-dual theory with locality shown in all orders was developed \cite{Didenko:2022qga}, moreover all these vertices were recently found explicitly \cite{Didenko:2024zpd} (see also \cite{Sharapov:2022nps} which is based upon \cite{Ponomarev:2016lrm}). Later, based on the system for the  (anti)holomorphic vertices, the off-shell generating system for HS fields in $d>4$ was developed \cite{Didenko:2023vna}. In the both generating systems all order locality was obtained due to the peculiar form of the star product
\begin{multline}\label{star_infty}
(f\ast g)(z,y,\bar{y})=\\
= \int du\,dv \, dP\, dQ\,d\bar{u}\, d\bar{v} e^{iu_\alpha v^\alpha -iP_\alpha v^\alpha+iu_\alpha Q^\alpha+i\bar{u}_{\dot{\alpha}} \bar{v}^{\dot{\alpha}}}f(z+u,y+P,\bar{y}+\bar{u})g(z+v,y+Q,\bar{y}+\bar{v}).
\end{multline} 
presented in \cite{Sharapov:2022faa}, where the same problem of obtaining spin-local vertices of the (anti)holomorphic sector of the 4d HS theory was considered. Since the product for the barred variables is the conventional Moyal-product we will not write the corresponding arguments and integrations explicitly in what follows for brevity. If both function are $z$-independent then \eqref{star_infty} is simply the Moyal-product  \eqref{Moyal} with respect to $y$ and $\bar{y}$ variables. Even though the product \eqref{star_infty} was not written explicitly in \cite{Didenko:2019xzz}, it can be obtained as the limit $\beta \rightarrow -\infty$ from the product written in the section devoted to re-orderings which after some regrouping and stretching casts into
\begin{multline}\label{star_beta}
(f \ast_\beta g)(z,y)=\int du\,dv \, dP\, dQ\, \frac{(1-\beta)^4}{\beta^2(2-\beta)^2}\exp\Bigg\{-\frac{i(1-\beta)^2}{\beta(2-\beta)} u_\alpha v^\alpha+\frac{i(1-\beta)^2}{\beta(2-\beta)} P_\alpha v^\alpha-\\-\frac{i(1-\beta)^2}{\beta(2-\beta)} u_\alpha Q^\alpha +\frac{i}{\beta(2-\beta)} P_\alpha Q^\alpha\Bigg\} f\Bigg(\frac{1-\beta}{1-\beta}(z+u),y+P\Bigg)g\Bigg(\frac{1-\beta}{1-\beta}(z+v),y+Q\Bigg)\,.
\end{multline}
For that reason the product \eqref{star_infty} will be called {\it the limiting star product} in what follows. Besides the obvious benefits it possess certain shortcomings, for example two one-forms in auxiliary $z$-direction  from  a proper functional class (to be defined later) cannot be multiplied since one faces uncertainties of the type $0\times \infty$ \footnote{One can recall that product \eqref{star_infty} comes from \eqref{star_beta}. The latter product does not bring any uncertainties, however considering product of three functions and carefully taking limit $\beta\rightarrow -\infty$ one faces non-associativity. I.e. $\lim_{\beta_1\rightarrow -\infty}\lim_{\beta_2\rightarrow -\infty}f\ast_{\beta_1} g\ast_{\beta_2}h\neq\lim_{\beta_2\rightarrow -\infty}\lim_{\beta_1\rightarrow -\infty}f\ast_{\beta_1} g\ast_{\beta_2}h$.}. This fact leads to the absence of the Leibniz rule for the differential in the auxiliary variables $z$ and thus the differential can be defined in various ways\footnote{This ambiguity might be related to the regularization procedure used in \cite{Sharapov:2022faa} as the part of the so-called {\it duality map}. Unfortunately rigorous justification for the regularization procedure is not available.}. Moreover it is no longer a differential in common sense but just an odd linear operator. 

We can consider the subset equations of the generating system proposed \cite{Didenko:2022qga} where differential does not adhere to the Leibniz rule. These equations have the following form analogous to \eqref{generate}
\begin{equation}\label{generate2}
\hat{A} \Phi^a_n=F^a(\Phi^b_{n-1},\Phi^c_{n-2},\Phi^d_{n-3},\ldots)\,.
\end{equation}
Here, $\hat{A}$ is an odd linear operator responsible for the $z$-dependence of the master fields. Even though in the particular choice made in \cite{Didenko:2022qga} it contains derivatives we deliberately choose this notation to emphasize that it is no longer a differential but just a linear map. Its definition cannot be inherited uniquely due to the violation of the Leibniz rule; hence, it must be defined in one way or another. Even after constraining its definition with the requirements:
\begin{equation}\label{consequence}
\hat{A} \hat{A}\equiv 0\,, \;\; \dr_x \hat{A}+\hat{A} \dr_x\equiv 0\, ,
\end{equation}
which simply implies that $\hat{A}$ is an odd linear operator, there are still numerous ways to define $\hat{A}$. We can inspect the implications of \eqref{consequence} on the equations of motion that were verified in the original paper and see that none of them are violated for any choice of $\hat{A}$. This naively suggests that there is no particular preference for the choice of $\hat{A}$ made in \cite{Didenko:2022qga} compared to other options. However, as we demonstrate explicitly below, this does not imply that the generating system produces $\dr_x$-consistent vertices for any choice of $\hat{A}$. The essence of the $\dr_x$-inconsistency of the resulting vertices lies in the expressions of the form:
\begin{equation}\label{intro}
(\dr_x \hat{A}+\hat{A}\dr_x)(\Phi^a_n\ast \Phi^b_m)\,.
\end{equation}
It was not shown in the original paper that expressions like this vanish on the equations of the generating system. Moreover, the absence of the Leibniz rule for $\hat{A}$ prevents the use of equations that describe evolution in the  variable $z$ to demonstrate that such expressions vanish; thus, another technique must be developed to resolve these expressions. Nonetheless, the generating system presented in \cite{Didenko:2022qga} is perfectly consistent in its original form when using the specific choice of $\hat{A}$. Proof of this statement is the main result of the current paper. It is performed by inspecting $\dr_x$-consistency of the vertices which result from the generating system. Analysing consistency one is able to confine oneself to specific ordering of the fields $\omega$ and $C$ since fields are allowed to take values in arbitrary associative algebra, in other words, by inspecting consistency the expressions with ordering $\go\go C\ldots C$, for example, and $\go C\go C\ldots C$ should vanish on their own. Considering the ordering $\go C\go C\ldots C$ we demonstrate that consistency rests on the highly nontrivial interplay between the shifted homotopies \cite{Gelfond:2018vmi}, \cite{Didenko:2018fgx} and the  limiting star product \eqref{star_infty}.

The paper is organized as follows. In section \ref{Leibniz}, we consider a toy example of a possible redefinition of the <<differential>> when the Leibniz rule is not available. In section \ref{BurnOriginal}, we recall the generating system for the (anti)holomorphic vertices originally presented in \cite{Didenko:2022qga} and show how interaction vertices are extracted, pointing out the gap in the proof of consistency. Then, in section \ref{BurnMango}, we consider a slightly twisted version of the original generating system, i.e., we utilize the freedom for redefining the <<differential>> explored in section \ref{Leibniz}. We demonstrate that an unfortunate choice of the differential might lead to inconsistent vertices. In the following section \ref{Proof}, we present a direct proof of consistency for the system proposed in \cite{Didenko:2022qga} by checking the $\dr_x$-consistency of the resulting vertices. In section \ref{Shifts}, we demonstrate that all the vertices that arise from the considered generating system can be expressed through the shifted homotopy formalism. The connection with the (anti)holomorphic sector of the 4d Vasiliev theory is discussed in section \ref{Vas_con}. Conclusion contains  discussion of the obtained results as well as yet open questions. In Appendices A and B, we present proof for the star-exchange-like identity for the limiting star-product and the generalization of the projection identity. Appendix C contains an explicit demonstration of the inconsistency arising from an unfortunate choice of the <<differential>>.

\section{Absence of the Leibniz rule and consequences}\label{Leibniz}
Consider the space of analytic functions of two one dimensional variables $z$ and $x$, denoted by $\Lambda^{0,0}$, along with the corresponding differential forms denoted as $\Lambda^{i,j}$, where $i$ specifies the degree in $\dr x$ and $j$ represents the degree in $\dr z$. One can choose the following basis in these spaces:
\begin{equation}
z^m x^n\,,\;\; \dr x\, z^m x^n\,,\;\; \dr z\, z^mx^n\,,\;\; \dr x\wedge \dr z\, z^m x^n\,.
\end{equation}
Here $m$ and $n$ are powers, not indices. The product for zero forms in $\dr z$ and $\dr x$ is conventional, i.e.
\begin{equation}\label{product}
z^{m_1} x^{n_1} \cdot z^{m_2} x^{m_2}:=z^{m_1+m_2}x^{n_1+n_2}\,.
\end{equation}
The same product is defined for zero-forms in $\dr z$ between one- and zero-forms in $\dr x$, i.e. for functions from $\Lambda^{0,0}$ and $\Lambda^{0,1}$. The differential $\dr_x$ is always given by the conventional expression
\begin{equation}
\dr_x :\Lambda^{0,0}\rightarrow \Lambda^{0,1},\;\; \dr_x:\Lambda^{1,0}\rightarrow \Lambda^{1,1},\;\; \dr_x:=\dr x\frac{\partial}{\partial x}.
\end{equation}
Defined in such a way,  $\dr_x$ obviously satisfies the Leibniz rule for the product \eqref{product}. However, here we assume that there is no product between zero-forms and one-forms in $\dr z$. In this sense one shouldn't care about the Leibniz rule for the differential $\dr_z$ and one can simply define it as a linear map from the space of zero-forms in $\dr z$ to the space of one-forms in $\dr z$. We define this map as follows
\begin{equation}
\dr_z: \Lambda^{0,0}\rightarrow \Lambda^{1,0},\;\; \dr_z (z^m x^n):=\dr z\, f_{m,n}(z) x^n.
\end{equation}
Here $f_{m,n}(z)$ is an arbitrary analytic function of $z$.

\begin{figure}[h]
\begin{center}
\begin{tikzpicture}
\filldraw[blue!50] (2,6) ellipse (1.5 and 0.7);
\filldraw[blue!50] (8,6) ellipse (1.5 and 0.7);
\filldraw[blue!50] (2,3) ellipse (1.5 and 0.7);
\filldraw[blue!50] (8,3) ellipse (1.5 and 0.7);
\node[below] at (8,3-0.7) {$\Lambda^{0,1}$};
\node[below] at (2,3-0.7) {$\Lambda^{0,0}$};
\node[above] at (2,6+0.7) {$\Lambda^{1,0}$};
\node[above] at (8,6+0.7) {$\Lambda^{1,1}$};
\filldraw (1.5,3) ellipse (0.02 and 0.02);
\filldraw (2,6.3) ellipse (0.02 and 0.02);
\filldraw (8,6.0) ellipse (0.02 and 0.02);
\filldraw (7,3.0) ellipse (0.02 and 0.02);
\draw[->,thick] (1.5,3) -- (2,6.3) ;
\draw[->,thick,dashed] (2,6.3) -- (8,6.0) ;
\draw[->,thick,dashed] (1.5,3) -- (7,3);
\draw[->,thick] (7,3) -- (8,6);
\node[left] at (1.7,4.5) {$\mathrm{d}_z$};
\node[above] at (5,3) {$\mathrm{d}_x$};
\node[above] at (5,6.2) {$\mathrm{d}_x$};
\node[left] at (7.5,4.5) {$\mathrm{d}_z$};
\end{tikzpicture}
\end{center}
\caption{Freedom in definition $\dr_z$.} \label{fig:fig1}
\end{figure}
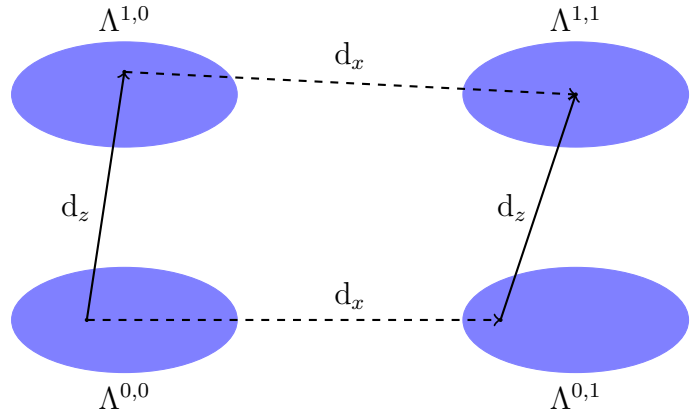

On the space of one-forms in $\dr x$ (labelled as $\Lambda^{0,1}$ in fig. \ref{fig:fig1}) we define  the differential as
\begin{equation}
\dr_z: \Lambda^{0,1}\rightarrow \Lambda^{1,1},\;\; \dr_z (\dr x z^m x^{n-1}):=\dr z\wedge \dr x\, f_{m,n}(z) x^{n-1}.
\end{equation}
Now we would like to check that the differentials defined this way anticommute on arbitrary analytic function. For this purpose consider the action of the differentials on the arbitrary basis element
\begin{equation}
\dr_z \, \dr_x\, (z^mx^n)=\dr_z(\dr x\, nz^mx^{n-1})=\dr z\wedge \dr x\, nf_{m,n}(z)x^{n-1}.
\end{equation}
Reverse order gives
\begin{equation}
\dr_x\, \dr_z\, (z^m x^n)=\dr_x(\dr z\, f_{m,n}(z) x^n)=\dr x\wedge \dr z\, nf_{m,n}(z) x^{n-1}.
\end{equation}
Hence the differentials defined in these ways anticommute on arbitrary analytic function in $z$ and $x$
\begin{equation}
(\dr_z\, \dr_x+\dr_x \, \dr_z)F(z,x)=0.
\end{equation}

This particular choice for $\dr_z$ is the illustration of the more general freedom than can be described by the commutative diagram shown on figure \ref{fig:fig1}. This diagram should be understood as follows. Pick  any\footnote{Working with generic spaces one should be careful with elements that are in kernel of $\dr_x$ within $\Lambda^{0,0}$. To maintain diagram commutative they should be mapped to elements of the kernel $\dr_x$ within $\Lambda^{1,0}$. Besides that, map from $\Lambda^{0,0}$ to $\Lambda^{1,0}$ can be indeed arbitrary.} element from $\Lambda^{0,0}$ and choose any linear map  such that its image is in $\Lambda^{1,0}$. Then in a unique way $\dr_x$ maps our original element and its image in $\Lambda^{1,0}$ to elements in $\Lambda^{0,1}$ and $\Lambda^{1,1}$ respectively. No matter how the original map $\dr_z : \Lambda^{0,0}\rightarrow \Lambda^{1,0}$ is chosen there is always a map $\dr_z :\Lambda^{0,1}\rightarrow \Lambda^{1,1}$ that makes diagram commutative. Moreover, if $\dr_x:\Lambda^{0,0}\rightarrow \Lambda^{0,1}$ is not the map onto the full space $\Lambda^{0,1}$ then one is free to define $\dr_z:\Lambda^{0,1}\rightarrow \Lambda^{1,1}$ on the quotient $\Lambda^{0,1}/{\dr_x(\Lambda^{0,0})}$ in an arbitrary way. This particular freedom will be used in the section \ref{BurnMango}. 

If the domain of $\dr_z$ and $\dr_x$ is bigger than just analytic functions, for example one may consider functions valued in some algebra (or depending on some additional variables), then there is even more freedom in the definition of $\dr_z$. For functions of the type $F^i(z,x)t_i$ we can draw multilayered diagrams like that in Fig.\ref{fig:fig1}, where each layer corresponds to the basis element $t_i$ of the algebra. It worth to stress that, when defining $\dr_z$, one is not confined to remain within the same layer. 

\section{Generating system for (anti)holomorphic sector of HS theory}\label{BurnOriginal}
The generating system proposed in \cite{Didenko:2022qga} looks as follows
\begin{align}
&\dr_{x} W+W*W=0\,,\label{dxWeq1}\\
&\dr_z W+\{W,\Lambda\}_{*}+\dr_x\Lambda=0\,,\label{Weq11}\\
&\dr_z\Lambda=C*\gga\,,\;\; \gamma:=\half\theta_\alpha\theta^\alpha e^{iz_\alpha y^\alpha}\,,\label{Lambdaeq1}\\
&\dr_x C*\gga=\dr_z\{W,\Lambda\}_{*}\,,\label{Ceq1}\\
&W\in \mathbf{C}^0\, , \label{class1}\\
&\Lambda[C]:=\theta^\alpha z_\alpha\int_0^1 d\tau \tau \, e^{i\tau z_\alpha y^\alpha}C(-\tau z|x)\, , \label{LambdaDef}
\end{align}
where product is given by \eqref{star_infty} and $\dr_z:=\theta^\alpha \frac{\partial}{\partial z^\alpha}$.  Various classes of functions denoted as $\mathbf{C}^r$  where $r=0,1,2$ indicates the degree in $\theta$ can be defined as follows. Consider the following generating expression
\begin{equation}\label{Classes}
\int \mathscr{D}\rho\int_0^1 d\mathcal{T}\, (1-\mathcal{T})^{1-r}\mathcal{T}^{r-1}\,\exp\{i\mathcal{T}z_\alpha (y-B)^\alpha+i(1-\mathcal{T})y^\alpha A_\alpha-i\mathcal{T}B_\alpha A^\alpha\}\,.
\end{equation}
Here $\int \mathscr{D}\rho$ is the shorthand notation for all the differentials of various $\rho$s and the measure of integrations, i.e.
\begin{equation}\label{D}
\int \mathscr{D}\rho:=\int \ldots \int d\rho_1\ldots d\rho_n\, \mu(\rho_1,\ldots \rho_n).
\end{equation}
The measure $\mu(\rho_1,\ldots,\rho_n)$ contains theta- and delta-functions of $\rho$s which makes the domain of integration a compact subset in $\mathbb{R}^n$. $A$ and $B$  are $\rho$-dependent linear combinations of various derivatives over full (anti)holomorphic spinorial arguments of the fields $\go$ and $C$ which are denoted as   
\begin{equation}\label{PT}
p_\alpha:=-i\frac{\partial}{\partial \mathsf{y}_C^\alpha}\, ,\;\; t_\alpha:=-i\frac{\partial}{\partial \mathsf{y}_\go^\alpha}\,\,.
\end{equation}
In that perspective the exponential  should be understood as follows
\begin{multline}
\exp\{i\mathcal{T}z_\alpha (y-B(t,p_1,p_2,\ldots))^\alpha+i(1-\mathcal{T})y^\alpha A(t,p_1,p_2,\ldots)_\alpha-\\-i\mathcal{T}B(t,p_1,p_2,\ldots)_\alpha A(t,p_1,p_2,\ldots)^\alpha\}\go(\mathsf{y}_\go|x)C(\mathsf{y}_{C_1}|x)C(\mathsf{y}_{C_2}|x)\ldots \Bigg|_{\mathsf{y}_\go=0,\; \mathsf{y}_{C_1}=0,\; \mathsf{y}_{C_2}=0,\ldots}
\end{multline}
A particular element from class $\mathbf{C}^r$ is obtained by taking the derivatives with respect to $A$ and $B$ of the generating expression \eqref{Classes}. Such differentiation brings additional powers of $\mathcal{T}$ and cancels the potentially dangerous poles. Only expressions that are free from divergences in $\mathcal{T}$ are the elements of the corresponding classes. The linear space property of classes defined in such a way rests on the corresponding properties of the integration measures (for the indirect proof, see Section 3 of \cite{Didenko:2022eso}).  As an illustration consider field $\Lambda$. One can easily bring the expression for $\Lambda$ given by \eqref{LambdaDef} to the form \eqref{Classes}
\begin{equation}
\Lambda[C]=-i\theta^\alpha\frac{\partial}{\partial {p}^\alpha}\int_0^1 dt\, e^{it z_\alpha (y+{p})^\alpha} C(\mathsf{y}_C|x)\Big|_{\mathsf{y}_C=0}\,.
\end{equation}
Here $A=0$ and $B=-p$ and according to \eqref{Classes} $\Lambda\in \mathbf{C}^1$.
In the original paper \cite{Didenko:2022qga} it is shown that the canonical differential in $z$ maps classes properly, i.e.
\begin{equation}\label{d_map}
\dr_z :\mathbf{C}^r\rightarrow \mathbf{C}^{r+1}\, 
\end{equation}
and products such as
\begin{equation}
\mathbf{C}^0 \ast \mathbf{C}^0\subset\mathbf{C}^0\, ,\;\;\; \mathbf{C}^0\ast \mathbf{C}^1\subset\mathbf{C}^1\, ,\;\;\; \mathbf{C}^1 \ast \mathbf{C}^0\subset\mathbf{C}^1
\end{equation}
are well defined and preserve grading. However the products 
\begin{equation}\label{miniTrouble}
\mathbf{C}^1\ast\mathbf{C}^1\,,\;\;\; \mathbf{C}^0\ast \mathbf{C}^2 
\end{equation}
are ill defined (one either faces the uncertainties like $\infty\times 0$ or the result of computation simply diverges) and hence one is not able to use the Leibniz rule acting with $\dr_z$ on expressions of the form $\mathbf{C}^0 \ast \mathbf{C}^1$ or $\mathbf{C}^1 \ast \mathbf{C}^0$.

Before we proceed with the consistency check we are going to demonstrate how this system {\it generates} equations of motions for (anti)holomorphic sector of the 4d HS theory, or, in other words, how one gets the equations like \eqref{sch0form},\eqref{sch1form} from \eqref{dxWeq1}-\eqref{LambdaDef}. One solves \eqref{Weq11} perturbatively (in powers of $C$)  and then substitutes the obtained results in \eqref{dxWeq1} and \eqref{Ceq1} to obtain equations on $C$ and $\go$.

Solving \eqref{Weq11} for $W$ in the zeroth order of $C$ we simply obtain that it is an arbitrary $z$-independent function which will be referred as $\omega$. Plugging it to \eqref{dxWeq1} and \eqref{Ceq1}, one respectively obtains
\begin{equation}\label{zero_curv}
\dr_x \omega+\omega \ast \omega+\mathcal{O}(C)=0\,,
\end{equation}
\begin{equation}\label{cov_const}
\dr_x C\ast \gamma=(-\omega \ast C+C\ast \pi[\omega]+\mathcal{O}(C^2))\ast \gamma\,\,\; \Longleftrightarrow\;\; \dr_x C+\omega \ast C-C\ast \pi[\omega]+\mathcal{O}(C^2)=0\,.
\end{equation}
Here $\pi$ is the automorphism of the $\ast$-algebra defined as
\begin{equation}\label{automorphism}
\pi[\Gamma(z,y)]:=\Gamma(-z,-y)\,.
\end{equation}
The equations \eqref{zero_curv},\eqref{cov_const} are familiar and obviously consistent. Solving for higher orders one obtains various nonlinear corrections (in $C$) to \eqref{zero_curv},\eqref{cov_const} (see \eqref{sch1form},\eqref{sch0form}). To obtain first correction to the field $W$  one solves \eqref{Weq11} which for this particular order of the perturbation casts into
\begin{equation}\label{dzWwC}
\dr_z W_{\go C}+\dr_z W_{C\go}+\{\go,\Lambda\}+\dr_x \Lambda\Big|_{\go C}+\dr_x \Lambda\Big|_{C\go}=0\, .
\end{equation} 
To make the notations clear we would like to note that one solves equations for particular orderings of $\omega$ and $C$ since system \eqref{dxWeq1}-\eqref{LambdaDef} is consistent for master fields taking values in arbitrary associative algebra just like the Vasiliev system \cite{Vasiliev:1999ba}. So, for the first order in $C$ the equation \eqref{Weq11} can be separated in two: one for the ordering $\go C$ and the other one for ordering $C\go$.  One might obtain a particular solution by virtue of the Poincaré lemma 
\begin{equation}
W_{\go C}=-\hmt_0 (\go \ast \Lambda)\, ,\;\; W_{C \go}=-\hmt_0 (C\ast \go)\,,
\end{equation}
where $\hmt_0$ operator is defined as follows:
\begin{equation}\label{Poincare}
\hmt_0(f(z,y|\theta)):=z^\alpha \frac{\partial}{\partial \theta^\alpha}\int_0^1 \frac{dt}{t}f(tz,y|t\theta)\,.
\end{equation}
The terms $\hmt_0(\dr_x \Lambda)$ vanish kinematically due to the specific form of $\Lambda$ \eqref{LambdaDef}. Plugging these results into \eqref{dxWeq1} and \eqref{Ceq1} we obtain nonlinear corrections to \eqref{zero_curv}, \eqref{cov_const} which look as follows
\begin{equation}\label{COMS_Chiral}
\dr_x \go+\go\ast \go=\underbrace{-\dr_x W_{\go C} -\dr_x W_{C\go}-\go \ast W_{\go C}-\go \ast W_{C\go}-W_{\go C}\ast \go-W_{C\go}\ast \go}+\mathcal{O}(C^2)\, ,
\end{equation}
\begin{equation}\label{Cubic_Chiral}
\dr_x C \ast \gamma=(C\ast \pi[\go]-\go \ast C)\ast \gamma+\underbrace{\dr_z\big(W_{\go C} \ast \Lambda+W_{C\go}\ast \Lambda+\Lambda\ast W_{\go C}+\Lambda\ast W_{C\go}\big)}+\mathcal{O}(C^3)\,.
\end{equation}
It is shown in the original paper that the evolution equations on $z$ for $W$ field namely \eqref{Weq11} are imposed in such a way that the underbraced term in \eqref{COMS_Chiral} is $z$-independent and the underbraced term in \eqref{Cubic_Chiral} has the form $\Upsilon(y|x)\ast \gamma$. The latter is true due to the remarkable {\bf projective identity} which holds for arbitrary $W\in \mathbf{C}^0$
\begin{equation}\label{Proj1}
\dr_z\left(W\ast \Lambda\right)=-\left(\int \dr^2 u\, \dr^2 v \,e^{iu_\alpha v^\alpha}\, W(z,y+u)C(y+v)\right)\Bigg|_{z=-y}\ast \, \frac{1}{2} \theta_\alpha \theta^\alpha e^{izy}\,,
\end{equation}
\begin{equation}\label{Proj2}
\dr_z\left(\Lambda\ast W\right)=\left(\int \dr^2 u \, \dr^2 v\, e^{iu_\alpha v^\alpha}\, C(y+u) W(z,-y-v)\right)\Bigg|_{z=-y}\ast \frac{1}{2}\theta_\alpha \theta^\alpha e^{izy}\,. 
\end{equation}
Different signs are due to the fact that $W$ is a spacetime one-form, i.e $\{\theta^\alpha,\dr x^\mu\}=0$.

This particular order is a good example where a difference between self-dual HS theory and the full HS dynamics is already manifest. One of the important results of the free HS dynamics that allowed to formulate eventually the full nonlinear system is the central on-mass-shell theorem \cite{Vasiliev:1999ba}. This theorem is very roughly the reformulation of the Fronsdal equations on $AdS_4$ \cite{Fronsdal:1978rb} in the unfolded form. The one-form sector of the unfolded version for the Fronsdal equations reads
\begin{equation}\label{OMS}
\dr_x \omega +\{\Omega,\go\}_\ast=\frac{i}{4}\bar{H}^{\dot{\alpha}\dot{\alpha}}\frac{\partial^2}{\partial \bar{y}^{\dot{\alpha}}\,\partial \bar{y}^{\dot{\alpha}}} C(0,\bar{y}|x)+\frac{i}{4}H^{\alpha\alpha} \frac{\partial^2}{\partial y^\alpha \partial y^\alpha} C(y,0|x)\,.
\end{equation}
Here $\Omega$ is the $AdS_4$ background connection
\begin{equation}
\Omega=\frac{i}{4}(\omega_{\alpha \beta} y^\alpha y^\beta+\bar{\go}_{\dot{\alpha}\dot{\beta}} \bar{y}^{\dot{\alpha}} \bar{y}^{\dot{\beta}}+2 e_{\alpha \dot{\alpha}} y^\alpha \bar{y}^{\dot{\alpha}})
\end{equation}
and $\bar{H}^{\dot{\alpha} \dot{\alpha}}$, $H^{\alpha \alpha}$ are the basis spatial two-forms
\begin{equation}
\bar{H}^{\dot{\alpha}\dot{\alpha}}=\frac{1}{4} e_\beta {}^{\dot{\alpha}}\, e^{\beta \dot{\alpha}}\,,\;\; H^{\alpha\alpha}=\frac{1}{4} e^\alpha {}_{\dot{\beta}}\, e^{\alpha \dot{\beta}}\,.
\end{equation}
In the self-dual theory HS theory either  the first or the second term of \eqref{OMS} is present, while the full unconstrained dynamics of the Fronsdal fields requires both. 

All nonlinear corrections that follow from \eqref{dxWeq1}-\eqref{LambdaDef} were explicitly calculated in \cite{Didenko:2024zpd}. The only requirement for this correction is the consistency with $\dr_x^2\equiv 0$. This check is performed by inspecting all possible differential consequences of \eqref{dxWeq1}-\eqref{LambdaDef}. I.e. one checks that identities
\begin{equation}\label{idenities}
\dr_x^2\equiv 0\, ,\;\; \dr_z^2\equiv 0\, ,\;\; \dr_x \dr_z+\dr_z \dr_x\equiv 0
\end{equation} 
are consistent with equations of motion \eqref{dxWeq1}-\eqref{LambdaDef}. This means that acting, for example, twice with $\dr_x$ on $W$ and using the equations there are will be no additional constraints except for the already written ones. A highly nontrivial problem is to show that 
\begin{equation}\label{bigTrouble}
(\dr_x \dr_z +\dr_z \dr_x)\{W,\Lambda\}_\ast=0\,.
\end{equation} 
Technical complications emerge due to the absence of the Leibniz rule since the expression of the form $\dr_z W\ast \Lambda$ does not exist. Even though $\dr_z\{W,\Lambda\}_\ast$ is well defined due to \eqref{class1} and the projective identities \eqref{Proj1}, \eqref{Proj2}, it is not clear how one should incorporate the $\dr_z$-dynamics of the $W$ field \eqref{Weq11} into consideration (more on this issue in section \ref{Check}). This particular consistency condition \eqref{bigTrouble} is vital for the $\dr_x$-consistency of the resulting vertices in the zero-form sector that come from \eqref{Ceq1}
\begin{equation}\label{trouble}
\dr_x^2 C\ast \gamma=\dr_x \dr_z\{W,\Lambda\}_\ast\,.
\end{equation}
To prove that the r.h.s. of \eqref{trouble} is indeed zero following the original paper \cite{Didenko:2022qga} we use {\it anticommutativity of differentials} $\dr_x$ and $\dr_z$, equations of motion and the property \eqref{class1}. Consider the different order of differentials on the r.h.s. of \eqref{trouble} 
\begin{multline}\label{Proof}
-\dr_z\dr_x\{W,\Lambda\}_\ast=-\dr_z\big(\dr_x W \ast \Lambda-W\ast \dr_x \Lambda+\dr_x \Lambda\ast W-\Lambda\ast \dr_x W\big)=\\
=-\dr_z\big(-W\ast W\ast \Lambda-W\ast(-\dr_z W-W\ast\Lambda-\Lambda\ast W)+(-\dr_z W-W\ast \Lambda-\Lambda\ast W)\ast W+\Lambda\ast W\ast W\big)=\\
=\dr_z\, \dr_z (W\ast W)\,.
\end{multline}
We would like to stress that due to the absence of the Leibniz rule one is not able to show that $\dr_z \dr_z (W\ast W)$ vanishes by virtue of equations \eqref{dxWeq1}-\eqref{Ceq1}. Vanishing rests exclusively on property \eqref{class1}, in other words due to $\eqref{class1}$ the product $W\ast W$ is well defined and $\dr_z^2$ acts on elements of $\mathbf{C}^0$ trivially. Eq. \eqref{Proof} implies that if \eqref{bigTrouble} is indeed valid then vertices produced by the generating system are $\dr_x$-consistent. The use of anticommutativity for $\dr_x$ and $\dr_z$ was not justified in the original paper\footnote{Slava Didenko provided the way to show that $\dr_x$ and $\dr_z=\theta^\alpha \frac{\partial}{\partial z^\alpha}$ actually anticommute on-shell in private conversation. It differs from the one presented in the paper.}. One should in fact check that the equations of motions are designed in a way that this property is not violated. In other words the off-shell anticommutativity, i.e. on the functions that are not constrained by any equations, does not imply that any equations of motion are necessary consistent with it. 

In the next section we investigate the freedom of definition of the differential in one-forms in $\theta$. We demonstrate how speculations with its definition might lead to inconsistent vertices even though all the conditions checked in \cite{Didenko:2022qga} are fulfilled. This freedom is provided by the absence of the Leibniz rule which puts all odd linear operators on completely equal footing. This should be understood as follows: {\it a priori} one does not have any operator for the map $\dr_z:\mathbf{C}^1\rightarrow\mathbf{C}^2$ since the absence of the Leibniz rule does not allow to inherit its definition in a unique way from the map $\mathbf{C}^0\rightarrow\mathbf{C}^1$. Hence one has to define it as odd linear map in one way or another. There is no particular preferences for $\theta^\alpha\frac{\partial}{\partial z^\alpha}$ compared to other possibilities at the definition stage. However, being plugged into the system \eqref{dxWeq1}-\eqref{LambdaDef} some of them lead to $\dr_x$-inconsistent vertices.



\section{Twisted version of the generating system for (anti)holomorphic sector of HS theory}\label{BurnMango}
\begin{figure}[h]
\begin{center}
\begin{tikzpicture}
\filldraw[blue!50] (2,6) ellipse (1.5 and 0.7);
\filldraw[blue!50] (8,6) ellipse (1.5 and 0.7);
\filldraw[blue!50] (2,3) ellipse (1.5 and 0.7);
\filldraw[blue!50] (8,3) ellipse (1.5 and 0.7);
\filldraw[blue!50] (14,3) ellipse (1.5 and 0.7);
\filldraw[blue!50] (14,6) ellipse (1.5 and 0.7);
\filldraw[green!50] (7.5,3) ellipse (0.8 and 0.4);
\filldraw[green!50] (7.5,6) ellipse (0.8 and 0.4);
\node[below] at (8,3-0.7) {$\mathbf{C}^1\wedge \dr x^\mu$};
\node[below] at (14,3-0.7) {$\mathbf{C}^1\wedge \dr x^\mu\wedge \dr x^\nu$};
\node[below] at (2,3-0.7) {$\mathbf{C}^1$};
\node[above] at (2,6+0.7) {$\mathbf{C}^2$};
\node[above] at (8,6+0.7) {$\mathbf{C}^2\wedge \dr x^\mu$};
\node[above] at (14,6+0.7) {$\mathbf{C}^2\wedge \dr x^\mu \wedge \dr x^\nu$};
\draw[->,thick] (2,3.7) -- (2,5.3);
\draw[->,thick] (8,3.7) -- (8,5.3);
\draw[->,thick] (8+6,3.7) -- (8+6,5.3);
\draw[->,thick,dashed] (3.5,6) -- (6.5,6.0);
\draw[->,thick,dashed] (3.5,3) -- (6.5,3);
\draw[->,thick,dashed] (3.5+6,6) -- (6.5+6,6.0);
\draw[->,thick,dashed] (3.5+6,3) -- (6.5+6,3);
\draw[->,>=stealth,ultra thick,blue] (3.5+6+0.25,3-0.25) -- (6.5+6-0.25,3-0.25);
\draw[->,>=stealth, ultra thick, blue] (8+6+0.25,3.7+0.25) -- (8+6+0.25,5.3-0.25);
\draw[->,>=stealth, ultra thick,red] (3.5+6+0.25,6-0.25) -- (6.5+6-0.25,6.0-0.25);
\draw[->,>=stealth, ultra thick,red] (8+0.25,3.7+0.25) -- (8+0.25,5.3-0.25);
\node[left] at (2,4.5) {$\mathfrak{d}$};
\node[above] at (5,3) {$\mathrm{d}_x$};
\node[above] at (5+6,3) {$\mathrm{d}_x$};
\node[above] at (5,6.0) {$\mathrm{d}_x$};
\node[above] at (5+6,6.0) {$\mathrm{d}_x$};
\node[left] at (8,4.5) {$\mathfrak{d}$};
\node[left] at (8+6,4.5) {$\mathfrak{d}$};
\node[below] at (8,1.5) {$\dr_x \mathbf{C}^1$};
\node[above] at (8,7.25) {$\dr_x \mathbf{C}^2$};
\draw[->, >=stealth,thick,black] (7.5,1.25) .. controls (6.25,1.5) and (6.5,2.5) .. (7.5,3);
\draw[->, >=stealth,thick,black] (7.25,7.6) .. controls (6.5,7.25) and (6.4,6.5) .. (7.5,6);
\end{tikzpicture}
\end{center}
\caption{New <<differential>> $\mathfrak{d}$ for one-forms in $\theta$.} \label{fig:fig2}
\end{figure}
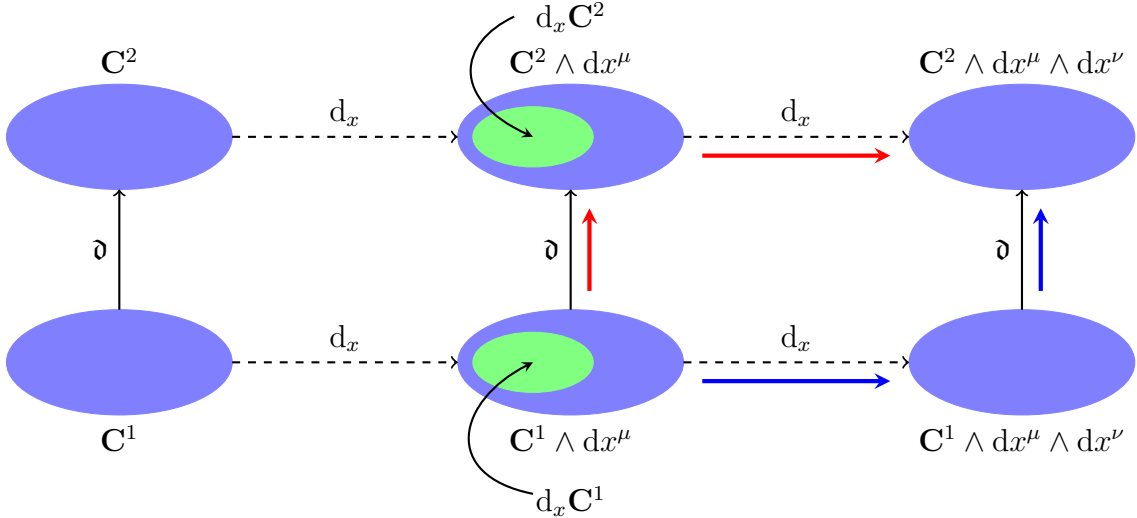

Suppose one changes the canonical $\dr_z:=\theta^\alpha \frac{\partial}{\partial z^\alpha}$ acting on one-forms in $\theta$ to $\mathfrak{d}$ that we are going to define through commutative diagram shown in Fig. \ref{fig:fig2}. We are going to consider now the twisted version of the original generating system that looks as follows
\begin{align}
&\dr_{x} W+W*W=0\,,\label{dxWeq}\\
&\dr_z W+\{W,\Lambda\}_{*}+\dr_x\Lambda=0\,,\label{Weq}\\
&\mathfrak{d}\Lambda=C*\gga\,,\label{Lambdaeq}\\
&\dr_x C*\gga=\mathfrak{d}\{W,\Lambda\}_{*}\,,\label{Ceq}\\
&W\in \mathbf{C}^0\, . \label{class}
\end{align}

As a preliminary step of defining the <<differential>> $\mathfrak{d}$ it is worth mentioning that the diagram shown in Fig. \ref{fig:fig2} has, in fact, an additional  level with functions from $\mathbf{C}^0$ class which is not depicted since no twists for the differential are available here (the Leibniz rule is valid for expressions of the form $\dr_z(\mathbf{C}^0\ast \mathbf{C}^0)$). The canonical differential $\dr_z$ maps them to the $\dr_z$-exact forms from the class $\mathbf{C}^1$ \eqref{d_map}. These exact forms form a subspaces, and in order to preserve nilpotency, i.e.,
\begin{equation}
\mathfrak{d}\, \dr_z=0
\end{equation}
we define $\mathfrak{d}$ on these subspaces in a trivial way, namely
\begin{equation}\label{trivial}
\mathfrak{d}\big|_{\dr_z \mathbf{C}^0}=\mathfrak{d}\big|_{\dr_z( \mathbf{C}^0\wedge \dr x^\mu)}=\mathfrak{d}\big|_{\dr_z(\mathbf{C}^0\wedge \dr x^\mu \wedge \dr x^\nu)}=0\,.
\end{equation}
Moreover we want only $\dr_z$-exact forms to be in the kernel of $\mathfrak{d}$. This requirement is not obligatory and serves demonstration purposes only. It allows to solve equations like
\begin{equation}
\dr_z f=g\,,\;\; f\in \mathbf{C}^0\,,\;\; g\in \mathbf{C}^1
\end{equation}
by virtue of Poincaré lemma for $\mathfrak{d}$-closed $g$. 

We define $\mathfrak{d}$ that maps $\mathbf{C}^1$ to $\mathbf{C}^2$ in a canonical way
\begin{equation}\label{C1_to_C2}
\mathfrak{d}:\mathbf{C}^1\rightarrow\mathbf{C}^2\,,\;\;\; \mathfrak{d}\big|_{\mathbf{C}^1}:=\theta^\alpha\frac{\partial}{\partial z^\alpha}\,.
\end{equation}
This particular definition implies that the equation \eqref{Lambdaeq} is fulfilled for the same $\Lambda$\footnote{Since $\Lambda$ is the only form that belongs to $\mathbf{C}^1$ we could define $\mathfrak{d}$ in a canonical way only for the subspace generated by $\eqref{LambdaDef}$ for all $C(y|x)$ with no constraints outside this subspace. }
\begin{equation}
\mathfrak{d}\big|_{\mathbf{C}^1}\left\{\theta^\alpha z_\alpha\int_0^1 d\tau \tau \, e^{i\tau z_\alpha y^\alpha}C(-\tau z|x) \right\}=C(y|x)\ast \gamma\,.
\end{equation}
This particular form of $\Lambda$ implies that \eqref{Weq} can be solved by virtue of Poincaré lemma since the source term of \eqref{Weq}, namely
$\{W,\Lambda\}_\ast+\dr_x \Lambda$, is $\mathfrak{d}$-closed due to \eqref{Ceq}. With this form of $\Lambda$ using the conventional homotopy operator \eqref{Poincare} one finds $W$ from the equation \eqref{Weq} in all orders in $C$ in a unique way. As the result in the unique way one finds the interaction vertices in the one-form sector \eqref{dxWeq1}.

To define $\mathfrak{d}$ on $\mathbf{C}\wedge \dr x^\mu$ we are going to use the freedom discussed in section \ref{Leibniz}. There is a subspace in $\mathbf{C}^1\wedge\dr x^\mu$ formed by the exact forms $\dr_x \mathbf{C}^1$ where no freedom is available, i.e. definition \eqref{C1_to_C2} obliges to define $\mathfrak{d}\big|_{\dr_x \mathbf{C}^1}$ in a unique way, namely 
\begin{equation}\label{C1_to_C2_add}
\mathfrak{d}\big|_{\dr_x \mathbf{C}^1}:=\theta^\alpha \frac{\partial}{\partial z^\alpha}\,,
\end{equation}
in order to maintain commutativity of the corresponding diagram. However on the quotient $\mathbf{C}^1 \wedge \dr x^\mu/\dr_x \mathbf{C}^1$ one is free to define $\mathfrak{d}$ in any way\footnote{Here we assume that the base manifold is topologically trivial, i.e. $H^1(\dr_x)=0$, and the bundle of $y$s and $z$s is glued to it as Cartesian product globally. However there are papers where nontrivial topology in $(Z,Y)$ space is considered \cite{DeFilippi:2019jqq}, \cite{Diaz:2024kpr}.}. In order to show the importance of this freedom recall how the generating system works when it comes to extracting vertices in the zero-form sector. To obtain them according to \eqref{Ceq} one acts with $\mathfrak{d}$ on $\{W,\Lambda\}$ (recall that due to the particular form of $\mathfrak{d}\big|_{\mathbf{C}^1}$ \eqref{C1_to_C2} $W$ is defined in a unique way if one applies conventional homotopy at each order of perturbation theory solving \eqref{Weq}). $\{W,\Lambda\}_\ast$ as an element of the vector space $\mathbf{C}^1\wedge \dr x^\mu$ has a unique decomposition in terms of
\begin{equation}
\mathbf{C}^1\wedge \dr x^\mu=\dr_x \mathbf{C}^1 \oplus \Big(\mathbf{C}^1 \wedge \dr x^\mu/\dr_x \mathbf{C}^1\Big)\,. 
\end{equation}
The procedure of taking this quotient is well defined since we are dealing with just vector spaces and one can treat elements of $\mathbf{C}^1$, $\mathbf{C}^1\wedge \dr x^\mu$, $\mathbf{C}^1\wedge \dr x^\mu\wedge \dr x^\nu$ as spacetime forms with additional color index. To find the result of the action of $\mathfrak{d}$ on $\{W,\Lambda\}_\ast$ we split $\{W,\Lambda\}_\ast$ into its exact part that belongs to $\dr_x \mathbf{C}^1$ and non-exact part (equivalence class) that belongs to the quotient $\mathbf{C}^1 \wedge \dr x^\mu/\dr_x \mathbf{C}^1$. Schematically the result of application of $\mathfrak{d}$ looks as follows
\begin{equation}
\mathfrak{d} \{W,\Lambda\}_\ast={\mathfrak{d}\big|_{\dr_x \mathbf{C}^1}(\text{exact part of } \{W,\Lambda\}_\ast)}+\mathfrak{d}\big|_{\mathbf{C}^1 \wedge \dr x^\mu/\dr_x \mathbf{C}^1}(\text{non-exact part of } \{W,\Lambda\}_\ast )\,.
\end{equation}
From the perspective of the resulting equations of motion (see \eqref{Ceq})
\begin{equation}\label{ku}
\dr_x C\ast \gamma=\underbrace{\mathfrak{d}\big|_{\dr_x \mathbf{C}^1}(\text{exact part of } \{W,\Lambda\}_\ast)}_{\dr_x F}+\mathfrak{d}\big|_{\mathbf{C}^1 \wedge \dr x^\mu/\dr_x \mathbf{C}^1}(\text{non-exact part of } \{W,\Lambda\}_\ast )
\end{equation}
the first term is trivial in the sense that it is exact off-shell, i.e. it is an exact form for any $\go(y|x)$ and $C(y|x)$. One can always get rid of such contributions by virtue of field redefinition. Moreover exact part is not even fixed by \eqref{C1_to_C2_add} since exact forms might come from the second term (see the r.h.s. of \eqref{ku}). The non-trivial part of the interaction vertices comes exclusively from action of $\mathfrak{d}$ on the non-exact part of $\{W,\Lambda\}_\ast$ which can be defined arbitrarily. Since the vertices in the one-form sector are already fixed in a unique way in all orders because of \eqref{C1_to_C2} there is simply no room for such tremendous freedom in the zero-form sector and additional constraints should be added to \eqref{dxWeq}-\eqref{class}. 

We would like to emphasize that one will not see any difference in solving the equations of \eqref{dxWeq1}-\eqref{LambdaDef} or \eqref{dxWeq}-\eqref{class} up to the point when one needs to compute $\mathfrak{d}\{W,\Lambda\}_\ast$ 
I.e. one perturbatively solves the equation responsible for $z$ dependence of $W$, namely  \eqref{Weq11}, with  the same source, since the expression for $\Lambda$ does not change due to \eqref{C1_to_C2} and hence the term $\hmt_0 (\dr_x \Lambda)$ vanishes. We would like to stress that by virtue of conventional homotopy the solution for the $W$ in any order is obtained for arbitrary $\go(y|x)$ and $C(y|x)$, so that $\{W,\Lambda\}_\ast$ is the generic form belonging to $\mathbf{C}^1\wedge \dr x^\mu$. If the system \eqref{dxWeq1}-\eqref{LambdaDef} produces nontrivial vertices (and it does. See \eqref{cov_const}), i.e. that are not $\dr_x$-exact off-shell, then by varying the definition of $\mathfrak{d}$ one might obtain any expressions for the nontrivial part. 

To make the problem even more evident consider the subspace within $\mathbf{C}^2\wedge \dr x^\mu$ of the functions of the form
\begin{equation}
\dr x^\mu f_\mu(y|x)\ast \gamma\,
\end{equation} 
and take a quotient of $\mathbf{C}^2\wedge \dr x^\mu$ over this subspace. Then  we are able to define $\mathfrak{d}$ on the quotient $\mathbf{C}^1 \wedge \dr x^\mu/\dr_x \mathbf{C}^1$ in the way that the image is contained in the quotient $\mathbf{C}^2 \wedge \dr x^\mu/\dr x^\mu f_\mu (y|x)\ast \gamma$
\begin{equation}\label{HugeTrouble}
\mathfrak{d}:\mathbf{C}^1 \wedge \dr x^\mu/\dr_x \mathbf{C}^1\rightarrow \mathbf{C}^2\wedge \dr x^\mu\;\; \text{such that}\;\; \mathfrak{d}\big(\mathbf{C}^1 \wedge \dr x^\mu/\dr_x \mathbf{C}^1\big)\subseteq \mathbf{C}^2 \wedge \dr x^\mu/\dr x^\mu f_\mu (y|x)\ast \gamma\,.
\end{equation} 
In this case the r.h.s. \eqref{Ceq} does not have the form $\dr x^\mu \Upsilon_\mu (y|x)\ast \gamma$  any more and hence in addition to \eqref{dxWeq}-\eqref{class} one should add new constraints to avoid contradiction, since the field $C$ is no longer $z$-independent. On the other hand just like in section \ref{BurnOriginal} we can consider a different order of $\dr_x$ and $\mathfrak{d}$ acting on $\{W,\Lambda\}_\ast$ and <<prove>>  the consistency
\begin{equation}
\mathfrak{d}\,\dr_x\{W,\Lambda\}=-\mathfrak{d}\,\dr_z(W\ast W)=0\,.
\end{equation}  

As an explicit demonstration for the above discussion, consider the following operator  on one-forms in $\theta$ 
\begin{equation}\label{definition}
\mathfrak{d}\Big[\theta^\sigma \Gamma_\sigma(z,y|x)\Big]:=\Bigg[i\int e^{iu_\alpha v^\alpha} \epsilon^{\alpha \xi}v_\alpha \frac{\partial}{\partial \theta^\xi} \theta^\sigma \Gamma_\sigma(-y+u,y+a|x)\Bigg]\ast \gamma\,,
\end{equation}
where $a$ is an arbitrary non-zero constant spinor. Defined in this way, $\mathfrak{d}$ obviously anticommutes with $\dr_x$:
\begin{equation}\label{Vital}
\dr_x\mathfrak{d}+\mathfrak{d}\dr_x\equiv0\,.
\end{equation}
After partial integration, the action of $\mathfrak{d}$ can be brought to the simpler form:
\begin{equation}
\mathfrak{d}\Big[\theta^\sigma \Gamma_\sigma(z,y|x)\Big]=\Big[\frac{\partial}{\partial u^\alpha}\Gamma^\alpha(u-y,y+a|x)\Big|_{u=0}\Big]\ast \gamma\,.
\end{equation}
There are one-forms in $\theta$ that, after the application of $\dr_z$, turn into $smth(y|x)\ast \gamma$, i.e., 
\begin{equation}\label{projectable}
\dr_z (\theta^\sigma \Gamma_\sigma(z,y|x))=\widetilde{\Gamma}(y|x)\ast \gamma
\end{equation}
with some $z$-independent $\widetilde{\Gamma}(y|x)$. We will call these one-forms {\it projectable}. For example, $\Lambda$ given by \eqref{LambdaDef} is one of these forms. By straightforward computations, one can easily show that on the space of projectable functions, $\dr_z$ and $\mathfrak{d}$ coincide:
\begin{equation}
\dr_z\Big|_{\text{projectable}}=\mathfrak{d}\Big|_{\text{projectable}}\,.
\end{equation}
$\dr_z$-exact forms are also projectable \eqref{projectable}, but with trivial $\widetilde{\Gamma}=0$, so property \eqref{trivial} is fulfilled. As is clear from definition \eqref{definition}, any function from $\mathbf{C}^1$ will have the form $smth(y|x)\ast \gamma$ after the application of $\mathfrak{d}$, which is not true for $\dr_z$. The latter makes $\mathfrak{d}$ even more preferable in this perspective. Indeed, one of the reasons why the original paper \cite{Didenko:2022qga} insisted on a rigid definition for $\Lambda$ as \eqref{LambdaDef} was the violation of the projection property \eqref{projectable} (see also \eqref{Proj1}, \eqref{Proj2}) if one chooses $\Lambda$ as
\begin{equation}\label{modified}
\Lambda'=\Lambda+\dr_z \varepsilon\,.
\end{equation}
for completely generic $\varepsilon\in \mathbf{C}^0$. In contrast, $\mathfrak{d}$ projects everything in the sense \eqref{projectable}. However, if one chooses $\mathfrak{d}$ as \eqref{definition} and selects a solution to \eqref{Lambdaeq} as \eqref{modified}, then already at the second order in $C$ one discovers inconsistency (see Appendix C for details). This is not surprising since condition \eqref{Vital} is violated on the equations of motion.

To summarize the observations made in this section, we would like to emphasize that commutative diagrams, such as shown in figure \ref{fig:fig2}, do not account for the imposed equations. Therefore, one must check that the equations of motion are {\it consistent} with the commutative diagram. The particular definition of $\mathfrak{d}$ given in \eqref{definition} clearly illustrates how this can be violated. In the case of the generating system presented in \cite{Didenko:2022qga}, only the path marked with blue arrows in fig. \ref{fig:fig2} was considered concerning the equations of motion. Since the absence of the Leibniz rule makes $\dr_z = \theta^\alpha \frac{\partial}{\partial z^\alpha}$ and $\mathfrak{d}$ indistinguishable, one is obligated to also study the red path because some of the $\mathfrak{d}$s can lead to inconsistent interactions, as pointed out in this section.

It is very unlikely that there is some other $\mathfrak{d}$s with \eqref{C1_to_C2} imposed that does not coincide with canonical $\dr_z=\theta^\alpha \frac{\partial}{\partial z^\alpha}$ leading nonetheless to consistent interaction. However if one relaxes \eqref{C1_to_C2} then there are $\mathfrak{d}$s such that \eqref{dxWeq}-\eqref{class} produces consistent vertices. Such $\mathfrak{d}$s should correspond to different field frames since it is very unlikely that there are two completely inequivalent versions of self-dual HS theory. 
At the current stage it is not clear how far in terms of field redefinitions one might get consistently varying the definition for $\mathfrak{d}$.

\section{Consistency check}\label{Check}
We are not going to address the problem of consistency of the system \eqref{dxWeq1}-\eqref{LambdaDef} in a standard way to avoid complications due to the absence of the Leibniz rule. The latter entails other complications; for example, inspecting $(\dr_z \dr_x+\dr_x \dr_z)W$ on the equations of motion yields\footnote{Here we have used the fact that $\dr_x \Lambda[C]=-\Lambda[\dr_x C]$ which is valid due to the definition \eqref{LambdaDef} and plugged r.h.s. of \eqref{Ceq1} as $\dr_x C$.}:
\begin{equation}\label{Light}
(\dr_z \dr_x+\dr_x\dr_z)W=-z^\alpha\frac{\partial}{\partial \theta^\alpha}\int_0^1 \frac{dt}{t}\Big(\dr_x\dr_z\{W,\Lambda\}_\ast\Big)(tz,y|t\theta)\,.
\end{equation}
This implies that one needs to keep track of usage equations describing evolution in $x$ and $z$ if one chooses to check the consistency following the path marked with red in Fig.\ref{fig:fig2}. To highlight the importance of \eqref{Light}, we would like to consider a somewhat deceptive approach to showing the equivalence of the paths marked in red and blue in Fig.\ref{fig:fig2} for the system \eqref{dxWeq1}-\eqref{class1} and point out the corresponding gap. We are going to compute $\dr_x \dr_z \{W,\Lambda\}_\ast$ by virtue of \eqref{Proj1}, \eqref{Proj2}:
\begin{multline}\label{naive1}
\dr_x\dr_z\{W,\Lambda\}_\ast=\int \dr^2 u\, \dr^2 v\, e^{iu_\alpha v^\alpha}\Big(\dr_x C(y+u)W(-y,-y-v)+\\+C(y+u)\dr_xW(-y,-y-v)-\dr_x W(-y,y+u)C(y+v)+W(-y,y+u)\dr_x C(y+v)\Big)\ast \gamma.
\end{multline}
Then we plug the $\dr_x$-dynamics for the $W$ and $C$ given by \eqref{dxWeq1},\eqref{Ceq1} respectively. For brevity we denote 
\begin{equation}
\mathcal{V}(W,C):=\int \dr^2 u\, \dr^2 v\, e^{iu_\alpha v^\alpha}\Big(C(y+u)W(-y,-y-v)-W(-y,y+u)C(y+v)\Big)\,,
\end{equation}
which allows to write dynamics in the zero-form sector, namely \eqref{Ceq1}, as
\begin{equation}
\dr_x C(y)=\mathcal{V}(W,C)(y)\,.
\end{equation}
Then \eqref{naive1} casts into
\begin{multline}\label{naive11}
\dr_x\dr_z\{W,\Lambda\}_\ast=\int \dr^2 u\, \dr^2 v\, e^{iu_\alpha v^\alpha}\Big(\mathcal{V}(W,C)(y+u) W(-y,-y-v)-\\-C(y+u)(W\ast W)(-y,-y-v)+(W\ast W)(-y,y+u)C(y+v)+\\+W(-y,y+u)\mathcal{V}(W,C)(y+v)\Big)\ast \gamma\,.
\end{multline}
On the other hand, we can consider $\dr_z \dr_x\{W,\Lambda\}_\ast$. Applying $\dr_x$, one has:
\begin{multline}
\dr_x \{W,\Lambda\}_\ast=\dr_x W\ast \Lambda[C]+W\ast \Lambda[\dr_x C]-\Lambda[\dr_x C]\ast W-\Lambda[C]\ast \dr_x W=\\
=-(W\ast W)\ast \Lambda[C]+W\ast\Lambda[\mathcal{V}(W,C)]-\Lambda[\mathcal{V}(W,C)]\ast W+\Lambda[C]\ast (W\ast W)\,.
\end{multline}
Then we apply $\dr_z$ to the previous expression, and using \eqref{Proj1}, \eqref{Proj2}, we obtain:
\begin{multline}\label{naive2}
\dr_z \dr_x\{W,\Lambda\}_\ast=-\int \dr^2 u\, \dr^2 v\, e^{iu_\alpha v^\alpha}\Big(\mathcal{V}(W,C)(y+u) W(-y,-y-v)-\\-C(y+u)(W\ast W)(-y,-y-v)+(W\ast W)(-y,y+u)C(y+v)+\\+W(-y,y+u)\mathcal{V}(W,C)(y+v)\Big)\ast \gamma\,.
\end{multline}
Naively, this implies that $\dr_x$ and $\dr_z$ anticommute on $\{W,\Lambda\}_\ast$ on the equations of the generating system. However, bringing the corresponding expressions to the form \eqref{naive11}, \eqref{naive2}, equation \eqref{Weq11} was completely eliminated from consideration. This makes the above consideration only {\it semi-on-shell}, since only $\dr_x$-dynamics was considered, identities \eqref{Proj1}, \eqref{Proj2} do not account for any equations describing evolution in $z$. In fact, one of these equations, namely \eqref{Weq11}, makes $\dr_z \dr_x \{W,\Lambda\}_\ast$ vanish (see \eqref{Proof}). 
From the perspective of \eqref{Light}, one cannot eliminate equation \eqref{Weq11} from consideration and then bring it back by demand; such manipulations may not be identical under the equations of motion. To highlight the importance of \eqref{Weq11} in this analysis, consider the system \eqref{dxWeq}-\eqref{class} and suppose one manages to come up with $\mathfrak{d}$ that has the following action on $\{W,\Lambda\}_\ast$:
\begin{equation}\label{Proj3}
\mathfrak{d}\left(W\ast \Lambda\right)=-\left(\int \dr^2 u\, \dr^2 v\, \dr^2 w \,K_L(u,v,w,y)\, W(w,y+u)C(y+v)\right)\ast \gamma\,,
\end{equation}
\begin{equation}\label{Proj4}
\mathfrak{d}\left(\Lambda\ast W\right)=\left(\int \dr^2 u \, \dr^2 v\, \dr^2 w\, K_R(u,v,w,y)\, C(y+u) W(w,-y-v)\right)\ast \gamma\,
\end{equation}
with some measures $K_{L,R}(u,v,w,y)$ and $\Lambda$-field still given by \eqref{LambdaDef}. Just as in the case with canonical $\dr_z$, one is able to show that $\{\dr_x, \mathfrak{d}\}$ vanish on $\{W,\Lambda\}_\ast$ (see \eqref{naive11},\eqref{naive2}). In fact, the only property required for the <<proof>> provided above is linearity in $W$ and $C$ of the expression $\mathfrak{d}\{W,\Lambda\}_\ast$ which is true for any linear operator $\mathfrak{d}$. However, as it is clear from the proceeding computations, the particular form of \eqref{Proj1}, \eqref{Proj2} is vital for the $\dr_x$-consistency of the vertices (see, for example, the proceeding section \ref{kinematic}). In other words, the result of applying $\dr_z$ (or some other $\mathfrak{d}$) should be consistent with the evolution equation in the variable $z$ imposed on the field $W$ otherwise vertices may be inconsistent with $\dr_x^2 \equiv 0$. Hence, the above semi-on-shell consideration does not imply consistency.

The approach proposed here is different and free complications that emerge from the absence of the Leibniz rule and its consequences \eqref{Light}. We are going to examine the $\dr_x$-consistency of the following system:
\begin{equation}\label{dxW69}
\dr_x W+W\ast W=0\,,
\end{equation}
\begin{equation}\label{dxC69}
\dr_x C=\int \dr^2 u\, \dr^2 v\, e^{iu_\alpha v^\alpha}\Big(C(y+u)W(-y,-y-v)-W(-y,y+u)C(y+v)\Big)\,.
\end{equation}
Here $W$ is some concrete functional of $\omega$ and $C$ given by
\begin{equation}\label{Ws}
W=\go(y|x)+W^{(1)}+W^{(2)}+\ldots\,,\;\; W^{(n)}=-\hmt_0\{W^{(n-1)},\Lambda\}_\ast\,,\;\; W^{(0)}=\omega(y|x)
\end{equation}
with $\Lambda$ given by \eqref{LambdaDef}.

We will prove the $\dr_x$-consistency of the system \eqref{dxW69},\eqref{dxC69} without using the fact that $W$ is a solution to any equations describing evolution in $z$.  Then we argue that $W$ can be treated as particular solution to equation
\begin{equation}\label{Weq69}
\dr_z W+\{W,\Lambda\}_\ast+\dr_x \Lambda=0\,.
\end{equation}
The latter eventually implies the consistency of the original system proposed in \cite{Didenko:2022qga}. 

We would like to point out the importance of \eqref{Weq69} from a different perspective.  Even after proving the $\dr_x$-consistency of \eqref{dxW69},\eqref{dxC69} this system is not describing (anti)holomorphic dynamics yet. The equation \eqref{dxW69} might be $z$ dependent. One can apply to \eqref{dxW69} projector $h_0$ which sets all $z$ variables to zero. Such projector obviously commutes with $\dr_x$ and hence equation
\begin{equation}\label{dxW68}
h_0\big(\dr_x W+W\ast W\big)=0\,
\end{equation}
is $\dr_x$-consistent, but system \eqref{dxW68},\eqref{dxC69} may be not in general, since this $z$-dependent part of \eqref{dxW69} can be vital for consistency of \eqref{dxC69}. However since $W$ are the solutions to \eqref{Weq69} then just as in the original paper \cite{Didenko:2022qga} one shows that \eqref{dxW69} and \eqref{dxW68} are equivalent.

Since \eqref{dxW69} is consistent for any $W$ we need to check only if \eqref{dxC69} is consistent with $\dr_x^2\equiv 0$. We do this at first for some particular orderings and then provide the proof for generic ordering with identities discovered on the previous steps.

The main issue that necessitates the use of unconventional methods to demonstrate consistency is that not all consequences derived from the system \eqref{dxWeq1}-\eqref{LambdaDef} are purely differential. This is due to the absence of the Leibniz rule for some operators. However, this does not diminish the value of the system; it simply makes the proof of consistency for the resulting vertices less straightforward than in the Vasiliev system.

\subsection{$\go \go C\ldots C$ consistency}\label{kinematic}
For brevity we introduce two <<products>> (see \eqref{Proj1}, \eqref{Proj2})
\begin{equation}\label{Ri}
W(z,y)\, \widehat{\ast}_R \,C(y):=\int \dr^2 u\, \dr^2 v\, e^{iu_\alpha v^\alpha}\, W(-y,y+u) \, C(y+v)\,,
\end{equation}
\begin{equation}\label{Le}
C(y)\, \widehat{\ast}_L\,  W(z,y) :=\int \dr^2 u\, \dr^2 v\, e^{iu_\alpha v^\alpha}\, C(y+u) W(-y,-y-v)\,. 
\end{equation}
We would like to stress that neither $\widehat{\ast}_R$ nor $\widehat{\ast}_L$ are associative. With this notation equation \eqref{Ceq1} casts into
\begin{equation}
\dr_x C=C \Le W-W\Ri C\,. 
\end{equation}
Consistency of this equation demands
\begin{multline}\label{(in)con}
\dr^2_x C\equiv0=\Big[(C\Le W)-(W\Ri C)\Big]\Le W-C \Le (W\ast W)+\\+(W\ast W)\Ri C+W\Ri\Big[(C\Le W)-(W\Ri C)\Big]
\end{multline}
where we have used the one-form dynamics $\dr_x W+W\ast W=0$. Note also that since the products are not associative computations should be performed in the particular order prescribed by  the brackets. 

Now we are going to examine \eqref{(in)con} and check if the structure $\omega^2 C^{n+1}$ vanish. Since the products used in the expression are not associative the $\dr_x$-consistency is not as clear as in Vasiliev system. Only two terms contribute to the required ordering, namely
\begin{equation}\label{ggCCC}
(W\ast W)\Ri C\Big|_{\omega^2 C^{n+1}}-W \Ri (W \Ri C)\Big|_{\omega^2 C^{n+1}}\,.
\end{equation}
We are going to examine if \eqref{ggCCC} is zero using generic form of $W_{\go C^n}$ from $\mathbf{C}^0$. We should check whether
\begin{equation}
(\omega \ast W_{\go C^n})\Ri C-\omega \Ri (W_{\go C^n} \Ri C)\, 
\end{equation} 
is identically zero.
Below we provide the results of necessary computations
\begin{multline}\label{wWCn}
\omega \ast W_{\go C^n}=\int \dr^2 u\, \dr^2 v\,  e^{iuv} \, \omega(y+u) W_{\go C^n}(z-v,y+v)=\\
=\int \dr^2 u \, \dr^2 v\,  e^{iuv} e^{i(y+u)^\alpha \widetilde{t}_\alpha}\omega W_{\go C^n}(z-v,y+v)=e^{iy^\alpha \widetilde{t}_\alpha }\omega W_{\go C^n}(z-\widetilde{t},y+\widetilde{t})\,,
\end{multline}
where $\widetilde{t}$ is derivative over the full (anti)holomorphic spinorial argument of the first $\go$-field in the expression (cf \eqref{PT}).
\begin{multline}\label{One}
(\omega \ast W_{\go C^n})\Ri C=\int \dr^2 u \, \dr^2 v\,  e^{iuv}\, e^{i(y+u)^\alpha \widetilde{t}_\alpha} \omega W_{\go C^n}(-y-\widetilde{t},y+u+\widetilde{t}) e^{i(y+v)^\alpha \widehat{p}_\alpha} C=\\
=e^{i(y-\widehat{p})^\alpha \widetilde{t}_\alpha+iy^\alpha \widehat{p}_\alpha}\omega W_{\go C^n}(-y-\widetilde{t},y-\widehat{p}+\widetilde{t})C\,,
\end{multline}
where $\widehat{p}$ is the derivative over the full (anti)holomorphic spinorial argument of the last $C$-field.

On the other hand, we have to compute
\begin{equation}
W_{\go C^n} \Ri C=\int \dr^2 u\, \dr^2 v\, e^{iuv} W_{\go C^n}(-y,y+u) e^{i(y+v)^\alpha \widehat{p}_\alpha}C=e^{iy^\alpha \widehat{p}_\alpha}W_{\go C^n}(-y,y-\widehat{p})C\,.
\end{equation}
\begin{multline}\label{Two}
\omega \Ri (W_{\go C^n}\Ri C)=\int \dr^2 u\, \dr^2 v\, e^{iuv}\, e^{i(y+u)^\alpha \widetilde{t}_\alpha} \omega e^{i(y+v)^\alpha \widehat{p}_\alpha}W_{\go C^n}(-y-v,y+v-\widehat{p})C=\\
=e^{iy^\alpha \widetilde{t}_\alpha} e^{i(y+\widetilde{t})^\alpha \widehat{p}_\alpha} \omega W_{\go C^n}(-y-\widetilde{t},y+\widetilde{t}-\widehat{p})C\,.
\end{multline}
Comparing \eqref{One} and \eqref{Two} one immediately sees that they coincide and hence this structures vanish in the consistency check. Moreover, we have shown that
\begin{equation}\label{ggCCC_}
(\go\ast W)\Ri C\Big|_{\omega (\omega,C,\ldots,C) C}-\go \Ri (W \Ri C)\Big|_{\omega (\omega,C,\ldots,C) C}=0\,.
\end{equation}
for any ordering $(\omega,C,\ldots, C)$ in the $W$-field.

Before we proceed to the proof for different ordering which is not that kinematically obvious as ordering $\go \go C\ldots C$ we recall the formalism of the shifted homotopies introduced in \cite{Gelfond:2018vmi} and then developed in \cite{Didenko:2018fgx}. Ironically being developed during the analysis of the Vasiliev generating system shifted homotopy formalism  breaks after the lowest nonlinear order, i.e. one cannot extract spin-local vertices from the Vasiliev generating system within the shifted homotopy approach beyond order $\mathcal{O}(C^2)$ of nonlinearity. In contrast, the formalism of the shifted homotopies is general enough to express all the vertices (modulo field redefinitions) that come from systems \cite{Didenko:2022qga}, \cite{Didenko:2023vna}.

\subsection{Shifted homotopies}
\label{R_shifted}
Operator of the shifted homotpy is defined as follows
\begin{equation}\label{h_def}
\hmt_q J(z,y|\theta):=(z+q)^\alpha \frac{\partial}{\partial \theta^\alpha}\int_0^1 \frac{dt}{t}\, J(tz-(1-t)q,y|t\theta)\, ,
\end{equation}
where $q$ is arbitrary $z$-independent spinor which also can be an operator. Such homotopy operators allow one to find particular solutions to equations of the form
\begin{equation}
\dr_z \Phi=J\,.
\end{equation}
for $\dr_z$-closed $J$. Particular solution can be written as 
\begin{equation}
\Phi=\hmt_q J
\end{equation}
which is indeed a solution due to the resolution of identity
\begin{equation}\label{unity}
\dr_z \hmt_q+\hmt_q \dr_z=1-h_q\,.
\end{equation}
Here $h_q$ is the projector to $\dr_z$-cohomology defined as
\begin{equation}\label{z-proj}
h_q f(z,y|\theta):=f(-q,y|0)\,.
\end{equation}
From the definition \eqref{h_def} and resolution of identity \eqref{unity} one can deduce the following  identities used in what follows \cite{Didenko:2018fgx}
\begin{equation}
\hmt_q \hmt_r=-\hmt_r \hmt_q\,,
\end{equation}
\begin{equation}
[\dr_z,\hmt_q\hmt_r]=\hmt_r-\hmt_q-h_q \hmt_r\,.
\end{equation}
What makes shifted homotopies special are the {\bf star-exchange} relations. These relations were originally developed for star product given as
\begin{equation}\label{Vasiliev_pr}
(f\ast^\prime g)(z,y)=\int {\dr^2 u\, \dr^2 v}\, e^{iu_\alpha v^\alpha}\, f(z+u,y+u)g(z-v,y+v)\,,
\end{equation}
which is different from the product \eqref{star_infty}. However for either $z$-independent $f$ or $g$ both these products give the same result. For the $z$-independent function $f$ and $y$-independent shift $q$ the following relations holds
\begin{equation}\label{SE_first}
\hmt_{q-p_f}\big(f(y)\ast \Gamma(z,y)\big)=f(y)\ast \hmt_q \Gamma(z,y)\, ,
\end{equation}
\begin{equation}\label{star_xchg}
\hmt_q \big(f(y)\ast \Gamma(z,y)\big)=f(y)\ast \hmt_{q+p_f}\Gamma(z,y)\,,
\end{equation}
\begin{equation}
\hmt_q\big(\Gamma(z,y)\ast f(y)\big)=\hmt_{q+p_f}\Gamma(z,y)\ast f(y)\,,
\end{equation}
\begin{equation}\label{SE_one}
\hmt_{q-p_f}\big(\Gamma(z,y)\ast f(y)\big)=\hmt_q \Gamma(z,y)\ast f(y)\,.
\end{equation}
Here $p_f:=-i\frac{\partial}{\partial y^\alpha}$ is the derivative over the full spinorial argument of function $f$ (recall \eqref{PT}). One can also consider an arbitrary dependence on $y$ for the shift parameter and the corresponding star-exchange relations  were deduced in \cite{Didenko:2018fgx}, but for the proceeding analysis only the following are in use
\begin{equation}
\hmt_{y+q}\big(f(y)\ast \Gamma(z,y)\big)=f(y)\ast \hmt_{y+q}\Gamma(z,y)\,,
\end{equation}
\begin{equation}
\hmt_{y-p_f}\big(\Gamma(z,y)\ast f(y)\big)=\hmt_{y+p_f}\Gamma(z,y)\ast f(y)\,.
\end{equation}
\begin{equation}
\hmt_{-y-p_f}\big(f(y)\ast \Gamma(z,y)\big)=f(y)\ast \hmt_{-y+p_f}\Gamma(z,y)\,.
\end{equation}
\begin{equation}\label{SE_last}
\hmt_{q-y}\big(\Gamma(z,y)\ast f(y)\big)=\hmt_{q-y}\Gamma(z,y)\ast f(y)\,.
\end{equation}
They hold for $y$-independent $q$. The same, i.e. one should simply replace $\hmt$ with $h$ with the same indices, set of star-exchange identities can be deduced for cohomology projectors. If one substitutes $\Gamma(z,y)$ with $\gamma$ defined as \eqref{Lambdaeq1} which possesses Klein-like properties
\begin{equation}
C(y)\ast e^{iz_\alpha y^\alpha}=C(-z)e^{iz_\alpha y^\alpha}\,,\;\; e^{iz_\alpha y^\alpha}\ast C(y)=C(z)e^{iz_\alpha y^\alpha}\,,
\end{equation}
\begin{equation}\label{klein_like}
e^{iz_\alpha y^\alpha} \ast C(y)=C(z) e^{iz_\alpha y^\alpha}=C(-y)\ast e^{iz_\alpha y^\alpha}\,.
\end{equation}

An immediate application of the shifted homotopy formalism is to the projection identity \eqref{Proj1},\eqref{Proj2}. Indeed, straightforward computation yields
\begin{equation}\label{proj_left}
\left(\int{\dr^2 u\, \dr^2 v }\, e^{iu_\alpha v^\alpha}\, C(y+u) W(z,-y-v)\right)\Bigg|_{z=-y}=h_{-y-\widehat{p}}\Big\{C(y)\ast\pi\big(W(z,y)\big)\Big\}\,
\end{equation}
\begin{equation}\label{proj_right}
\left(\int {\dr^2 u\, \dr^2 v}\, e^{iu_\alpha v^\alpha}\, W(z,y+u)C(y+v)\right)\Bigg|_{z=-y}=h_{y-\widehat{p}}\Big\{W(z,y)\ast C(y)\Big\}\,.
\end{equation}
Here $\widehat{p}$ stands for the derivative with respect to the full spinorial argument of the $C$-field written explicitly and $\pi$ is automorphism \eqref{automorphism}. So, the equation \eqref{dxC69} turns into
\begin{equation}
\dr_x C=h_{-y-\widehat{p}}\Big\{C(y)\ast\pi\big(W(z,y)\big)\Big\}-h_{y-\widehat{p}}\Big\{W(z,y)\ast C(y)\Big\}
\end{equation}
or with the help of star-exchange identities to the form
\begin{equation}
\dr_x C=C(y)\ast h_{-y+\widehat{p}}\Big\{\pi\big(W(z,y)\big)\Big\}-h_{y+\widehat{p}}\Big\{W(z,y)\Big\}\ast C(y)\,.
\end{equation}
Consistency of the previous equation implies that the following holds due to equation imposed on $C$ and $W$
\begin{multline}\label{Consistency}
0=h_{-y-\widehat{p}}\Big\{C(y)\ast\pi\big[W(z,y)\big]\Big\}\ast h_{-y+\widehat{q}}\Big(\pi\big[W(z,y)\big]\Big)-h_{-y-\widehat{p}}\Big\{C(y)\ast \pi\big[W(z,y)\ast W(z,y)\big]\Big\}-\\
-h_{y-\widehat{p}}\Big\{W(z,y)\ast C(y)\Big\}\ast h_{-y+\widehat{q}}\Big(\pi\big[W(z,y)\big]\Big)+h_{y-\widehat{p}}\Big\{W(z,y)\ast W(z,y)\ast C(y)\Big\}+\\
+h_{y+\widehat{q}}\Big(W(z,y)\Big)\ast h_{-y-\widehat{p}}\Big\{C(y)\ast \pi\big[W(z,y)\big]\Big\}-h_{y+\widehat{q}}\Big(W(z,y)\Big)\ast h_{y-\widehat{p}}\Big\{W(z,y)\ast C(y)\Big\}\,.
\end{multline}
Here $\widehat{p}$ is the derivative of the only $C$-filed written explicitly and $\widehat{q}$ is the derivative over $y$ of the full another argument of star-product multiplication. The latter should be understood as follows. Consider, for example, the term
\begin{equation}\label{q_example}
h_{y+\widehat{q}}\Big(W(z,y)\Big)\ast h_{-y-\widehat{p}}\Big\{C(y)\ast \pi\big[W(z,y)\big]\Big\}\,.
\end{equation}
One can straightforwardly way compute the right factor
\begin{equation}
\mathcal{A}(y)=h_{-y-\widehat{p}}\Big\{C(y)\ast \pi\big[W(z,y)\big]\Big\}\,.
\end{equation}
Then \eqref{q_example} is treated as
\begin{equation}\label{Q_example}
h_{y+\widehat{q}}\Big(W(z,y)\Big)\ast h_{-y-\widehat{p}}\Big\{C(y)\ast \pi\big[W(z,y)\big]\Big\}=\Big(h_{y-i\frac{\partial}{\partial \mathsf{y}_{\mathcal{A}}}}\Big(W(z,y)\Big)\ast\mathcal{A}(y+\mathsf{y}_{\mathcal{A}})\Big)\Big|_{\mathsf{y}_{\mathcal{A}}=0}\,.
\end{equation}
Formula \eqref{Consistency} provides simple a proof for the consistency for ordering $\go\go C^N$ for any $N\geq 1$. Indeed
\begin{equation}
\dr_x^2 C\Big|_{\go\go C^N}=h_{y-\widehat{p}}\big\{W\ast W\ast C\big\}\Big|_{\go \go C^N}-h_{y+\widehat{q}}(W)\ast h_{y-\widehat{p}}(W\ast C)\Big|_{\go\go C^N}\,,
\end{equation}
other structures do not contribute to this particular ordering of fields. Moreover, only specific $W$s, namely  $\go$ and $W_{\go C^{N-1}}$, contribute to this ordering and hence one gets
\begin{multline}\label{kin_vanish}
\dr_x^2 C\Big|_{\go\go C^N}=h_{y-\widehat{p}}\big\{\go \ast W_{\go C^{N-1}}\ast C\big\}-h_{y+\widehat{q}}(\go)\ast h_{y-\widehat{p}}(W_{\go C^{N-1}}\ast C)=\\
=h_{y-\widehat{p}}\big\{\go \ast W_{\go C^{N-1}}\ast C\big\}-\go\ast h_{y-\widehat{p}}(W_{\go C^{N-1}}\ast C)=\\
=h_{y-\widehat{p}}\big\{\go \ast W_{\go C^{N-1}}\ast C\big\}-h_{y-\widehat{p}}\big\{\go \ast W_{\go C^{N-1}}\ast C\big\}\equiv 0\,.
\end{multline}
Other orderings of $\omega$s and $C$s are not that kinemtatically obvious and rely on identities that provide interplay between shifted homotopies and limiting star product.

\subsection{$\go C\go C\ldots C$ consistency}\label{non_kinematic}
In this section we focus on the $\go C \go C\ldots C$ ordering and thus consistency requirement gives
\begin{multline}\label{goCgoC}
\dr_x^2 C\Big|_{\go C \go C\ldots C}=-h_{y-\widehat{p}}\Big\{\go(y)\ast C(y)\Big\}\ast h_{-y+\widehat{q}}\Big(\pi\big[W_{\go C\ldots C}(z,y)\big]\Big)+\\
+h_{y-\widehat{p}}\Big\{W_{\go C}(z,y)\ast W_{\go C\ldots C}(z,y)\ast C(y)\Big\}+\underline{h_{y-\widehat{p}}\Big\{\go(y)\ast W_{C\go C\ldots C}(z,y)\ast C(y)\Big\}}+\\
+h_{y+\widehat{q}}\Big(\go(y)\Big)\ast h_{-y-\widehat{p}}\Big\{C(y)\ast \pi\big[W_{\go C\ldots C}(z,y)\big]\Big\}-\\-
\underline{h_{y+\widehat{q}}\Big(\go(y)\Big)\ast h_{y-\widehat{p}}\Big\{W_{C\go C\ldots C}(z,y)\ast C(y)\Big\}}-h_{y+\widehat{q}}\Big(W_{\go C}(y)\Big)\ast h_{y-\widehat{p}}\Big\{W_{\go C\ldots C}(z,y)\ast C(y)\Big\}\,.
\end{multline}
With the help of the star-exchange identities one can show that the underlined terms trivially vanish (see \eqref{kin_vanish} and \eqref{ggCCC_}) and one is left with the four terms
\begin{multline}\label{goCgoCC}
\dr_x^2 C\Big|_{\go C \go C^N}=-h_{y-\widehat{p}}\Big\{\go(y)\ast C(y)\Big\}\ast h_{-y+\widehat{q}}\Big(\pi\big[W_{\go C^N}(z,y)\big]\Big)+\\
+h_{y-\widehat{p}}\Big\{W_{\go C}(z,y)\ast W_{\go C^{N-1}}(z,y)\ast C(y)\Big\}+\\
+h_{y+\widehat{q}}\Big(\go(y)\Big)\ast h_{-y-\widehat{p}}\Big\{C(y)\ast \pi\big[W_{\go C^N}(z,y)\big]\Big\}-\\-h_{y+\widehat{q}}\Big(W_{\go C}(y)\Big)\ast h_{y-\widehat{p}}\Big\{W_{\go C^{N-1}}(z,y)\ast C(y)\Big\}\,.
 \end{multline}
Here we specify the particular order in the perturbation theory for the fields $W$. From the representation \eqref{Ws} one can easily deduce the following representation for $W_{\go C^N}$
\begin{equation}\label{doll}
W_{\go C^N}=-\hmt_0\big(W_{\go C^{N-1}}\ast \Lambda\big)\,.
\end{equation}
The only explicit expression we are going to use in the proof of consistency is 
\begin{equation}\label{WwC}
W_{\go C}=\go\ast C \ast \hmt_{p+t}\hmt_p\gamma\,.
\end{equation}
With the help of the star-exchange relations and \eqref{doll} the r.h.s. of \eqref{goCgoCC} can be brought to the form
\begin{multline}\label{ku6}
\omega\ast C \ast (h_{-y+p_1+t_1}-h_{-y+p_1})\Big(\pi[W_{\go C^{N-1}}]\Big)+\omega\ast C\ast h_{y-p_{N+1}}(\hmt_{p_1+t_1}\hmt_{p_1+t_1}\gamma \ast W_{\go C^{N-1}}\ast C)-\\
-\omega\ast C\ast h_{y+\widehat{q}}\hmt_{p_1+t_1}\hmt_{p_1}\gamma \ast h_{y-p_{N+1}}\Big(W_{\go C^{N-1}}\ast C\Big)\,.
\end{multline}
The presence of $\widehat{q}$ in the index of the last term makes the last term of \eqref{ku6} somewhat special since the shifts of the other terms contain only derivatives of $\omega$s and $C$s\footnote{For $\go C\go C$ ordering, for example, $\widehat{q}$ should also contain derivative over $y$ of expression of the type $h_a \hmt_b \hmt_c \gamma$ \eqref{hddg} that comes from \eqref{WwC}. To handle this type of shifts within framework of the Vasiliev theory the limiting shifting homotopy was developed in \cite{Didenko:2019xzz}.}. However this unhandy $\widehat{q}$ can be eliminated by virtue of the star-exchange identities
\begin{multline}
-\omega\ast C\ast h_{y+\widehat{q}}\hmt_{p_1+t_1}\hmt_{p_1}\gamma \ast h_{y-p_{N+1}}\Big(W_{\go C^{N-1}}\ast C\Big)=\\
=-\omega\ast C\ast h_{\widehat{q}}\hmt_{-y+p_1+t_1}\hmt_{-y+p_1}\gamma \ast h_{y-p_{N+1}}\Big(W_{\go C^{N-1}}\ast C\Big)=\\
=-\omega \ast C \ast h_0\Big(\hmt_{-y+p_1+t_1}\hmt_{-y+p_1}\gamma \ast h_{y-p_{N+1}}\big(W_{\go C^{N-1}}\ast C\big)\Big)\,.
\end{multline}
Using star-exchange once again the r.h.s. of \eqref{goCgoCC} casts into
\begin{multline}\label{lbl1}
\omega \ast C\ast \Big\{(h_{-y+p_1+t_1}-h_{-y+p_1})\hmt_{p_{N+1}}\Big(\pi[W_{\go C^{N-1}}]\ast \hmt_{p_{N+1}}\gamma\Big)+h_{y+p_{N+1}}\big(\hmt_{p_1+t_1}\hmt_{p_1}\gamma \ast W_{\go C^{N-1}}\big)-\\
-h_{p_{N+1}}\Big(\hmt_{-y+p_1+t_1}\hmt_{-y+p_1}\gamma \ast h_{y+p_{N+1}}(W_{\go C^{N-1}}\Big)\Big\} \ast C\,.
\end{multline}
For brevity in what follows we use the following shorthand notation:
\begin{equation}
p_1+t_1=a\, ,\;\; p_1=b\,,\;\; p_{N+1}=c\,,\;\; W_{\go C^{N-1}}=W\,.
\end{equation}
Using the resolution of identity we work out the first term of \eqref{lbl1}
\begin{multline}
(h_{-y+a}-h_{-y+b})\hmt_c\big(\pi[W]\ast \hmt_c \gamma\big)=\\=\hmt_{y-b}\hmt_{y-a} \dr_z\big(\pi[W]\ast \hmt_c \gamma\big)-h_{-y+b}\hmt_{-y+a}\big(\pi[W]\ast \hmt_c \gamma\big)+\\+\hmt_c(\hmt_{-y+b}-\hmt_{-y+a})\dr_z \big(\pi[W]\ast \hmt_c\gamma\big)\,.
\end{multline}
Both sides are $z$-independent and applying the $h_c$ cohomology projector one obtains
\begin{multline}
(h_{-y+a}-h_{-y+b})\hmt_c\big(\pi[W]\ast \hmt_c \gamma\big)=\\=h_c\hmt_{y-b}\hmt_{y-a} \dr_z\big(\pi[W]\ast \hmt_c \gamma\big)-h_{-y+b}\hmt_{-y+a}\big(\pi[W]\ast \hmt_c \gamma\big)\,.
\end{multline}
To proceed with further simplifications  we use the following generalization of the projection identity (for the proof see Appendix B)
\begin{equation}\label{proj_new}
\dr_z\Big(\Gamma(z,y)\ast \hmt_c \gamma\Big)=h_{y-c}\big(\Gamma(z,y)\big)\ast \gamma\;\;\;\; \forall \Gamma(z,y)\in \mathbf{C}^0\,.
\end{equation}
\begin{multline}
(h_{-y+a}-h_{-y+b})\hmt_c\big(\pi[W]\ast \hmt_c \gamma\big)=\\=-h_c\hmt_{y-b}\hmt_{y-a}\Big(h_{y-c}\big(\pi[W]\big)\ast \gamma\Big)-h_{-y+b}\hmt_{-y+a}\big(\pi[W]\ast \hmt_c \gamma\big)\,.
\end{multline}
Note the sign change which appears since $W$ is a spacetime one-form. Using the fact that $h_{y-c}\big(\pi[W]\big)$ is $z$-dependent and the Klein-like properties of $\gamma$ \eqref{klein_like} one derives
\begin{multline}
(h_{-y+a}-h_{-y+b})\hmt_c\big(\pi[W]\ast \hmt_c \gamma\big)=\\=-h_c\big(\hmt_{y-b}\hmt_{y-a}\gamma\ast h_{y+c}W\big)-h_{-y+b}\hmt_{-y+a}\big(\pi[W]\ast \hmt_c \gamma\big)\,.
\end{multline}
Plugging simplified expressions back into \eqref{lbl1} we obtain
\begin{multline}
\omega \ast C\ast \Big\{-h_c\big(\hmt_{-y+b}\hmt_{-y+a}\gamma \ast h_{y+c}W\big)-h_{-y+b}\hmt_{-y+a}\big(\pi[W]\ast \hmt_c \gamma\big)+\\
+h_{y+c}(\hmt_a\hmt_b \gamma\ast W)-h_c(\hmt_{-y+a}\hmt_{-y+b}\gamma \ast h_{y+c} W)\Big\}\ast C\,.
\end{multline}
Here the remaining terms vanish due to the star-exchange-like identity for the limiting star product (for the proof see Appendix A)
\begin{equation}\label{star_new}
h_{y+c}(\hmt_a\hmt_b \gamma\ast \Gamma(z,y))=h_{-y+b}\hmt_{-y+a}\big(\pi[\Gamma(z,y)]\ast \hmt_c \gamma\big)\;\;\; \forall\Gamma(z,y)\in \mathbf{C}^0\,.
\end{equation}
We would like to emphasise the great difference with the star-exchange relations known before \eqref{SE_first}-\eqref{SE_last}, in \eqref{star_new} both of the multiplied functions are $z$-dependent.

\subsubsection*{Projective identities}
Identity \eqref{proj_new} and its counterpart
\begin{equation}\label{proj_new2}
\dr_z\big(\hmt_a \gamma \ast \Gamma(z,y)\big)=\gamma \ast h_{-y-a}\big(\Gamma(z,y)\big)=h_{-y+a}\big(\pi[\Gamma(z,y)]\big)\ast \gamma\;\;\; \forall\Gamma(z,y)\in \mathbf{C}^0
\end{equation}
allows us to specify class of functions $\Phi(z,y|\theta)\in\mathbf{C}^1$ that possess projective property
\begin{equation}\label{project}
\dr_z \big(\Gamma(z,y)\ast \Phi(z,y|\theta)\big)=smth(y)\ast \gamma\,,\;\;\; \dr_z\big(\Phi(z,y|\theta)\ast \Gamma(z,y)\big)=smth(y)\ast \gamma
\end{equation}
for arbitrary $\Gamma(z,y)\in \mathbf{C}^0$. One of those functions is $\Lambda$ (see \eqref{Proj1},\eqref{Proj2}), however identities \eqref{proj_new},\eqref{proj_new2} provide a wider class of candidates
\begin{equation}
f(z,y)\ast \hmt_a \gamma\, ,\;\; \hmt_a\gamma \ast f(z,y)
\end{equation}
for arbitrary $f(z,y)\in \mathbf{C}^0$ and $y$-independent shift $a$. However one should be careful with the order of multiplication since identities \eqref{proj_new},\eqref{proj_new2} do not cover this kind of expressions
\begin{equation}\label{proj_new4}
\dr_z\big(\Gamma(z,y)\ast\hmt_b \gamma \ast f(z,y)\big)\,.
\end{equation}
Nonetheless straightforward computation yields that such products are also projectable (see Appendix B for details). Unfortunately we are not able to provide simple expression for \eqref{proj_new4} like \eqref{proj_new} or \eqref{proj_new2}.

One of the reason for a rigid definition of $\Lambda$ \eqref{LambdaDef} in \cite{Didenko:2022qga} was the failure of the projection property if one adds arbitrary exact form to $\Lambda$. With the help of identities \eqref{proj_new},\eqref{proj_new2} and \eqref{proj_new3} we argue that the following exact forms do not violate projective properties
\begin{equation}\label{candide}
\dr_z \hmt_a\big(\Gamma \ast \hmt_b\gamma\big)\,, \;\; \dr_z \hmt_a\big(\hmt_b\gamma \ast \Gamma\big)
\end{equation}
for arbitrary $\Gamma\in\mathbf{C}^0$ and $y$-independent shifts $a$ and $b$. Indeed, consider for example the first candidate. By virtue of the resolution of identity \eqref{unity} and \eqref{proj_new} we have
\begin{multline}
\dr_z \hmt_a(\Gamma\ast \hmt_b \gamma)=\Gamma\ast\hmt_b \gamma-\hmt_a \dr_z(\Gamma \ast \hmt_b \gamma)=\Gamma \ast \hmt_b\gamma-\hmt_a\big(h_{y-b}\Gamma\ast \gamma\big)=\\=\Gamma \ast \hmt_b \gamma-h_{y-b}\Gamma\ast \hmt_{a+\widehat{q}}\gamma\,.
\end{multline}
Here $\widehat{q}$ is the derivative over the full holomorphic argument of $h_{y-b}\Gamma$ and also we use the star-exchange identity \eqref{star_xchg}. Now consider
\begin{equation}
\dr_z\big(\Phi\ast \dr_z\hmt_a (\Gamma\ast \hmt_b \gamma)\big)=h_{y-b}(\Phi\ast \Gamma)\ast \gamma-h_{y-a-\widehat{q}}\big(\Phi\ast h_{y-b}\Gamma\big)\ast \gamma\,.
\end{equation} 
This implies that one can adjust $\Lambda$ \eqref{LambdaDef} with a $\dr_z$-exact form of the type $\dr_z \hmt_a(\Gamma\ast \hmt_b \gamma)$ and the resulting $\Lambda$ still possess projective properties under $\dr_z$ application \eqref{project}. In similar way one shows that second candidate of \eqref{candide} is also valid.


\subsection{Generic Ordering}\label{Full_proof}
With the help of identity \eqref{proj_new} one can easily show consistency for the  generic ordering. We start from \eqref{Consistency} and multiply both sides by $\gamma$. By virtue of the star-exchange identities and \eqref{proj_new} we can rewrite for example the last term as follows
\begin{equation}
-h_{y+\widehat{q}}(W)\ast h_{y-\widehat{p}}(W\ast C)\ast \gamma=
-h_{y-\widehat{q}^\prime}(W)\ast \gamma \ast \pi\Big[h_{y-\widehat{p}}(W\ast C)\Big]\,.
\end{equation}
Here $\widehat{q}^\prime$ is the derivative over full antiholomorphic argument of $\pi\Big[h_{y-\widehat{p}}(W\ast C)\Big]$ (see \eqref{q_example}-\eqref{Q_example}). Moreover, we have used Klein-like properties of $\gamma$ \eqref{klein_like}. By virtue of \eqref{proj_new} it casts into
\begin{equation}
\dr_z\Big(W\ast \hmt_{\widehat{q}^\prime}\gamma\Big)\ast \pi\Big[h_{y-\widehat{p}}(W\ast C)\Big]=\dr_z\Big(W\ast \hmt_{\widehat{q}^\prime}\gamma\ast \pi\Big[h_{y-\widehat{p}}(W\ast C)\Big]\Big)\,.
\end{equation}
Even though the Leibniz rule is not available for $\dr_z$ in its entirety , the multiplication with $z$-independent functions does not exhibit any uncertainties, i.e.
\begin{equation}
\dr_z \big(\mathbf{C}^1 \big)\ast f(y)=\dr_z\big(\mathbf{C}^1 \ast f(y)\big)\,, \;\; f(y)\ast \dr_z \mathbf{C}^1=\dr_z\big(f(y)\ast \mathbf{C}^1\big)\, ,
\end{equation}
which can be shown by straightforward computations. To proceed we use \eqref{SE_one} and \eqref{proj_new} once again
\begin{multline}
\dr_z\Big(W\ast \hmt_0\Big(\gamma\ast \pi\Big[h_{-\widehat{p}}(W\ast C)\Big]\Big)\Big)=\dr_z \Big(W\ast \hmt_0\Big(h_{y-p}(W\ast C)\ast \gamma\Big)\Big)=\\
=-\dr_z\Big(W\ast \hmt_0 \dr_z\Big(W\ast C\ast \hmt_p \gamma\Big)\Big)=-\dr_z\Big(W\ast \hmt_0 \dr_z (W\ast \Lambda)\Big)\,.
\end{multline}
At the last step we use the definition of $\Lambda$ \eqref{LambdaDef} and its form in term of the shifted homotopies
\begin{equation}
\Lambda[C]=C\ast \hmt_p \gamma\,.
\end{equation}
For other terms we provide only final expressions
\begin{equation}
h_{-y-\widehat{p}}(C\ast \pi[W])\ast h_{-y+\widehat{q}}(\pi[W])\ast \gamma=\dr_z \Big(\hmt_0 \big(\dr_z (\Lambda\ast W)\big)\ast W\Big)\,,
\end{equation}
\begin{equation}
-h_{-y-\widehat{p}}(C\ast \pi[W\ast W])\ast \gamma=-\dr_z\big(\Lambda\ast W\ast W\big)\,,
\end{equation}
\begin{equation}
-h_{y-\widehat{p}}(W\ast C)\ast h_{-y+\widehat{q}}(\pi[W])\ast \gamma=\dr_z \Big(\hmt_0\big(\dr_z (W\ast\Lambda)\big)\ast W\Big)\,,
\end{equation}
\begin{equation}
h_{y-\widehat{p}}(W\ast W\ast C)\ast \gamma=\dr_z\big(W\ast W\ast \Lambda\big)\,,
\end{equation}
\begin{equation}
h_{y+\widehat{q}}(W)\ast h_{-y-\widehat{p}}(C\ast \pi[W])\ast \gamma=-\dr_z\Big(W\ast\hmt_0\big(\dr_z(\Lambda\ast W)\big)\Big)\,.
\end{equation}
Combining all the results together \eqref{Consistency} casts into
\begin{equation}
\dr_x^2 C\ast \gamma=\dr_z \Big(\hmt_0\big(\dr_z\{W,\Lambda\}_\ast\big)\ast W-\Lambda\ast W\ast W+W\ast W\ast \Lambda-W\ast \hmt_0\big(\dr_z\{W,\Lambda\}_\ast\big)\Big)\,.
\end{equation}
Resolution of identity for shifted homotopy \eqref{unity} yields
\begin{multline}
\dr_x^2 C\ast \gamma=\dr_z\Big((1-\dr_z \hmt_0-h_0)(\{W,\Lambda\}_\ast)\ast W-\Lambda\ast W\ast W+\\+W\ast W\ast\Lambda+W\ast W\ast \Lambda
-W\ast(1-\dr_z \hmt_0-h_0)\{W,\Lambda\}_\ast\Big)\,.
\end{multline}
Using the definition for $W$ \eqref{Ws} we obtain
\begin{equation}\label{proof} 
\dr_x^2 C\ast \gamma=\dr_z\Big(-\dr_z(W-\go)\ast W+W\ast\dr_z(W-\go)\Big)\,.
\end{equation}
From this form it is already obvious that the r.h.s. of \eqref{proof} vanishes identically. Even though one can notice similiarities with \eqref{Proof} there is a big difference. Bringing expression on the r.h.s. to the form $\dr_z \dr_z (W\ast W)$ no equation describing evolution in $z$ for $W$ were used, just \eqref{Ws}. Now when $\dr_x$-consistency of \eqref{dxW69} and \eqref{dxC69} is proven one can treat $W$ given by \eqref{Ws} as particular perturbative solution to \eqref{dxW68}
\begin{multline}
\dr_z W=\dr_z(\go-\hmt_0\{W,\Lambda\})=(-1+\hmt_0\dr_z)\{W,\Lambda\}_\ast=-\{W,\Lambda\}_\ast+\hmt_0\dr_x C\ast \gamma=\\
=-\{W,\Lambda\}_\ast-\dr_x \hmt_0 C\ast \gamma=-\{W,\Lambda\}_\ast-\dr_x \Lambda\,.
\end{multline} 
This finishes the proof of consistency of  the system \eqref{dxWeq1}-\eqref{LambdaDef}.

Even though this way of showing the consistency, i.e. all orders at the same time, is perfectly valid it does not reveal nontrivial relations like \eqref{star_new} which might be in great use for developing generalization of the system \eqref{dxWeq1}-\eqref{LambdaDef}.

\section{Shifted homotopies and (anti)holomorphic sector}\label{Shifts}
Shifted homotopies are not just the tool that provides to show consistency. In fact within this formalism one is able to express all the vertices that follow from system \eqref{dxWeq1}-\eqref{LambdaDef}.

To obtain $W$s one perturbatively solves \eqref{Weq11}. To make reulting vertices spin-local \eqref{Weq11} is solved by virtue of conventional homotopy, i.e. with zero shift. Below we show that all contributions to $W$ can be written in the compact form by virtue of shifted homotopy approach. For simplicity we confine ourselves to ordering $\go C\ldots C$. In the lowest order of perturbation theory $\mathcal{O}(C^0)$ we have simply $\go$. In the first order we have
\begin{equation}
\dr_z W_{\go C}+\go \ast \Lambda+\dr_x \Lambda\big|_{\go C}=0\,.
\end{equation}
Solving for $W_{\go C}$ by virtue of conventional homotopy, with trivial shift, one obtains
\begin{equation}
W_{\go C}=-\hmt_0\big(\go \ast \Lambda\big)\,.
\end{equation}
Using the star-exchange relations it can be brought to the factorized form
\begin{equation}
W_{\go C}=\hmt_0\big(\go \ast C \ast \hmt_p \gamma\big)=\go \ast C\ast \hmt_{p+t}\hmt_p \gamma\,.
\end{equation}
We can proceed further and solve \eqref{Weq11} for $W_{\go CC}$. Corresponding equation is
\begin{equation}
\dr_z W_{\go CC}+W_{\go C}\ast \Lambda+\dr_x \Lambda\big|_{\go CC}=0\,.
\end{equation}
Particular solution provided by conventional homotopy gives
\begin{multline}
W_{\go CC}=-\hmt_0\big(\go \ast C\ast \hmt_{p_1+t}\hmt_{p_1}\gamma \ast C\ast \hmt_{p_2}\gamma\big)=-\hmt_0\big(\go \ast C \ast C(-y)\ast \hmt_{p_1+t-2p_2}\hmt_{p_1-2p_2}\gamma \ast \hmt_{p_2}\gamma\big)=\\=\go\ast C\ast C(-y)\ast \hmt_{p_1+t-p_2}\big(\hmt_{p_1+t-2p_2}\hmt_{p_1-2p_2}\gamma \ast \hmt_{p_2}\gamma\big)\,.
\end{multline}
The sign flip in argument of the second $C$ appears due to Klein-like properties of $\gamma$ \eqref{klein_like}. It is pretty clear that with shifted homotopies one can write down a particular solution in any order of perturbation theory in the form
\begin{equation}
W_{\go C^N}=\go\ast C\ast C(-y)\ast C\ast C(-y)\ast \ldots C((-)^{N+1}y)\ast \hmt_{t+p_1-p_2+\dots+(-)^{N+1}p_N}\big(\ldots\big)\,.
\end{equation}
It is worth to mention that using purely shifted homotopies one obtains pretty much the same expressions for $W_{\go CC}$ solving equations of Vasiliev generating system with one imortant difference that emerges in second order and beyond. Products \eqref{star_infty} and \eqref{Vasiliev_pr} make difference. To produce local vertices within Vasiliev framework different homotopy approach was developed, which is not neccesary for the system \eqref{dxWeq1}-\eqref{LambdaDef}.

To obtain vertices in zero-forms we plug obtained $W_{\go C}$ and $W_{C\go}$ into \eqref{Ceq1}. The result looks as follows
\begin{equation}\label{first}
\Upsilon_{\go CC}=-\omega(y)\ast C(y)\ast C(-y)\ast h_{y-p_2} \hmt_{p_1+t-2p_2} \hmt_{p_1-2p_2}\gamma\,,
\end{equation}
\begin{equation}
\Upsilon_{CC\go}=C(y)\ast C(-y)\ast \go(y)\ast h_{-y+p_1-2p_2+2t}\hmt_{-p_2+t}\hmt_{-p_2+2t}\gamma\, ,
\end{equation}
\begin{multline}\label{last}
\Upsilon_{C\go C}=C(y)\ast \go(-y)\ast C(-y)\ast(h_{-y+p_1-2t-2p_2}\hmt_{-p_2-t}\hmt_{-p_2}\gamma- \\-h_{y-p_2}\hmt_{p_1-t-2p_2}\hmt_{p_1-2t-2p_2}\gamma)\,
\end{multline}
With $h_c \hmt_b \hmt_a \gamma$ is given by
\begin{equation}\label{hddg}
h_c \hmt_b\hmt_a \gamma=2\int d^3\tau_+\, \delta(1-\sum_{i=1} ^3 \tau_i)(b-c)_\alpha (a-c)^\alpha \exp\big\{-i(\tau_1 c+\tau_2 b+\tau_3 a)_\alpha y^\alpha\big\}\,
\end{equation}
which can be easily deduced from definitions \eqref{h_def}, \eqref{z-proj}, also the following shorthand notation was used
\begin{equation}
d^3 \tau_+:=\dr \tau_1\, \dr \tau_2\, \dr \tau_3\, \theta(\tau_1)\theta(\tau_2)\theta(\tau_3)\,.
\end{equation}

For one-forms, since equation \eqref{dxWeq1} is $z$-independent by virtue of \eqref{Weq11}, one can write down vertices in the $\mathcal{O}(C^1)$ order as
\begin{equation}
\Upsilon(\go,C,C)=h_0(-\dr_x W_{\go C} -\dr_x W_{C\go}-\go \ast W_{\go C}-\go \ast W_{C\go}-W_{\go C}\ast \go-W_{C\go}\ast \go)\,.
\end{equation}
Since $W_{\go C}$ and $W_{C\go}$ are obtained by virtue of conventional homotopy $\hmt_0$ contributions with $\dr_x$ are trivial
\begin{equation}
h_0(\dr_x W_{\go C})=h_0(\dr_x W_{C\go})=0\,.
\end{equation}
After application  of the star-exchange identities vertices  acquire the following compact form
\begin{equation}\label{firs1}
\Upsilon_{\go\go C}=-\go(y)\ast \go(y)\ast C(y)\ast h_{p+t_1+t_2}\hmt_{p+t_2}\hmt_p\gamma\,,
\end{equation}
\begin{multline}
\Upsilon_{\go C \go}=-\go(y)\ast C(y)\ast \go(-y)\ast h_{t_1+p-t_2}\hmt_{p-t_2}\hmt_{p-2t_2}\gamma-\\-\go(y)\ast C(y)\ast \go(-y)\ast h_{t_1+p-t_2}\hmt_{p_1+t_1-2t_2}\hmt_{p-2t_2}\gamma\,,
\end{multline}
\begin{equation}\label{last1}
\Upsilon_{C\go\go}=-C(y)\ast \go(-y)\ast \go(-y)\ast h_{p-t_1-t_2}\hmt_{p-t_1-2t_2}\hmt_{p-2t_1-2t_2}\gamma\,.
\end{equation}
This particular order is a simple demonstration but it is clear that all the vertices can be expressed by virtue of the shifted homotopy formalism since all the $W$ fields can be expressed in such a way. These factorized form like \eqref{first}-\eqref{last} and \eqref{firs1}-\eqref{last1} can be important for developing factorization procedure that puts $d$-dimensional system \cite{Didenko:2023vna} on-shell. Indeed, performed above the  proof of the  consistency shows that it rests on identities that involves structures like
\begin{equation}
h_{a_1}\hmt_{a_2}\hmt_{a_2}\gamma\,,\;\; h_{a_1}\hmt_{a_2}(\hmt_{a_3}\hmt_{a_4} \gamma \ast \hmt_{a_5} \gamma)\,,\;\; h_{a_1}\hmt_{a_2}\big(\hmt_{a_3}(\hmt_{a_4}\hmt_{a_5}\gamma\ast \hmt_{a_6}\gamma)\ast \hmt_{a_7}\gamma\big)\,,\ldots
\end{equation}
For example, checking consistency for ordering $\go C\go C$ one obtains
\begin{multline}
\go(y,\bar{y})\ast C(y,\bar{y})\ast \go(-y,\bar{y})\ast C(-y,\bar{y})\ast\Big\{-h_{-y+p_1+t_1-2p_2-2t_2}\hmt_{-p_2-t_2}\hmt_{-p_2}\gamma+\\+h_{-y+p_1-2p_2-2t_2}\hmt_{-p_2-t_2}\hmt_{-p_2}\gamma-h_{y-p_2-t_2}\hmt_{p_1+t_1-2p_2-2t_2}\hmt_{p_1-2p_2-2t_2}\gamma+\\+h_{y-p_2}\hmt_{p_1+t_1-2t_2-2p_2}\hmt_{p_1-2p_2-2t_2}\gamma\Big\}\,.
\end{multline}
As was shown in \cite{Didenko:2018fgx} expression in brackets vanishes identically. We deliberately reincarnate barred variables here. Particular form of the product for $\bar{y}$, namely \eqref{Moyal}, is not important, as long as it is associative vertices are $\dr_x$-consistent. The same semi-silent role auxiliary variables $\vec{\mathbf{y}}_\alpha^i$ with  $i=0,\ldots,d-1$ play in $d$-dimensional system \cite{Didenko:2023vna}. I.e. particular form of product for $\vec{\mathbf{y}}_\alpha^i$ used in \cite{Didenko:2023vna}, namely
\begin{equation}
f(\vec{\mathbf{y}})\,\vec{\star}\,g(\vec{\mathbf{y}}):=\int \dr^{2d}\vec{\mathbf{u}}\,\, \dr^{2d}\vec{\mathbf{v}}\,\, e^{i\vec{\mathbf{u}}_\alpha {}^i \vec{\mathbf{v}}^{\alpha j}\eta_{ij}}\,f(\vec{\mathbf{y}}+\vec{\mathbf{u}})g(\vec{\mathbf{y}}+\vec{\mathbf{v}})\,,
\end{equation}
is not relevant for $\dr_x$-consistency of the resulting vertices. Any associative modification describes consistent interaction.

\section{Connection with (anti)holomorphic sector of 4d Vasiliev HS theory}
\label{Vas_con}
A mechanism that relates 4d Vasiliev theory and \eqref{dxWeq1}-\eqref{LambdaDef} is an open and important question since besides all the advantages like all order spin-locality of the self-dual theory it is not clear how it should be consistently deformed in order to include the mixed sector which is present in Vasiliev theory from the very beginning. A yet more important question if such a deformation of the self-dual theory exists at all. One can address the problem  in the following way: one computes all the (anti)holomorphic vertices in Vasiliev theory and looks for field redefinition that might map them to the vertices that follow from self-dual system. This approach does not look technically promising even though some results on all order computations from Vasiliev system are available \cite{Didenko:2015cwv}. A more promising approach is to search for the map on the level of the generating system, i.e. to find proper reformulation/deformation of either \eqref{dxWeq1}-\eqref{LambdaDef} or (anti)holomorphic truncation of the Vasiliev theory.

The path to the interaction verticies in Vasiliev theory in the zero-form (anti)holomorphic sector is different from the ones in the systems  \cite{Didenko:2022qga} or \cite{Didenko:2023vna}. In Vasiliev generating system verticies appear as cohomologies of $\dr_z$ while in other systems as images of the $\dr_z$ (see \eqref{Ceq1}).  
\begin{figure}[h]
\begin{center}
\begin{tikzpicture}
\filldraw[blue!20] (8,3) ellipse (1.9 and 1.30);
\filldraw[blue!50] (8,6) ellipse (1.5 and 0.7);
\filldraw[blue!50] (8,3) ellipse (1.5 and 0.7);
\filldraw[green!50] (8,3) ellipse (1 and 0.4);
\filldraw[green!50] (8,6) ellipse (1 and 0.4);
\node[above] at (8,6+0.7) {$\mathbf{C}^2\wedge \dr x^\mu$};
\draw[->,thick] (8,3.7) -- (8,5.3);
\node[left] at (8,4.5) {$\mathfrak{d}$};
\node[left] at (6,3) {$\hmt_0(\Upsilon^{Vas}(y|x)\ast \gamma)$};
\draw[->,>=stealth] (6,3) -- (7.5,3);
\node[left] at (6,6) {$\Upsilon^{Vas}(y|x)\ast \gamma$};
\draw[->,>=stealth] (6,6) -- (7.5,6);
\draw[->,>=stealth, thick] (8.3,6) .. controls (9,4.8) and (8.7,4) .. (8.3,3);
\node[right] at (8.8,4.5) {$\hmt_0$};
\draw[->,>=stealth] (10.5,3) -- (9.7,3);
\node[left] at (5.8,2.3) {$\mathbf{C}^1 \wedge \dr x^\mu$};
\draw[->,>=stealth] (5.8,2.3) -- (8,2.45);
\node[right] at (10.5,3) {Gelfond-Vasiliev};
\node[right] at (10.5,2.5) {functional class $\mathcal{H}^1$};
\end{tikzpicture}
\end{center}
\caption{Connection with (anti)holomorphic Vasiliev vertices.} \label{fig:fig3}
\end{figure}
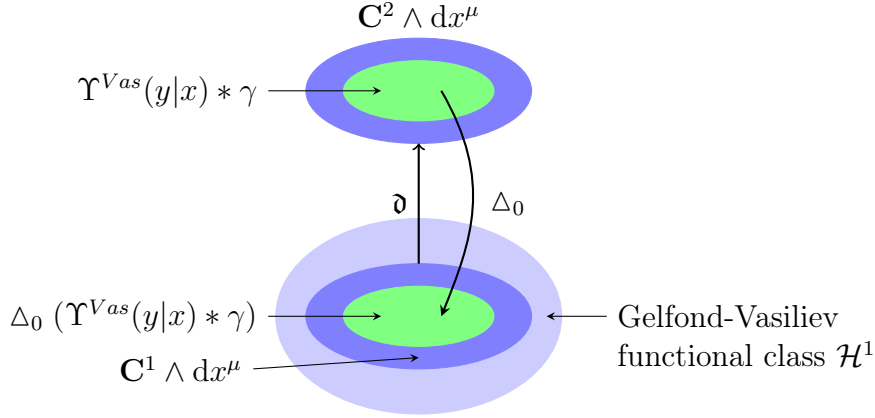
Even though the path is different resulting vertices still have the form of 
analytic in $y$ function with $x$-dependent coefficients $\Upsilon^{{Vas}}(y|x)$\footnote{This should be understand as follows. One picks up some functions $\omega(y|x)$, $C(y|x)$ and plugs them to vertices generated by Vasiliev system $\Upsilon(\omega,C,C)+\Upsilon(\go,C,C,C)+\ldots=\Upsilon^{{Vas}}(y|x)$}. This implies that the functions of the form $\Upsilon^{{Vas}}(y|x)\ast \gamma$ are still within space $\mathbf{C}^2\wedge \dr x^\mu$ (see Fig. \ref{fig:fig3}). One can in an injective and linear way map all such functions to $\mathbf{C}^1 \wedge \dr x^\mu$, for example by virtue of conventional homotopy operator
\begin{equation}
\hmt_0\big(\Upsilon^{Vas}(y|x)\ast \gamma\big)\in \mathbf{C}^1 \wedge \dr x^\mu\,.
\end{equation} 
This implies that the class $\mathbf{C}^1\wedge \dr x^\mu$ is large enough so that in general one can build a map $\mathfrak{f}:\mathbf{C}^1 \wedge \dr x^\mu\rightarrow \mathbf{C}^2 \wedge \dr x^\mu$ to these vertices. This observation allows one to bypass possible obstruction with much bigger Gelfond-Vasiliev functional class $\mathcal{H}^1$ defined in \cite{Gelfond:2019tac}. However this is not the end of the story. Suppose we want to obtain Vasiliev (anti)holomorphic vertices using twisted verion of the original system \eqref{dxWeq}-\eqref{class}. It is not clear if these vertices are reachable as $\mathfrak{d}\{W,\Lambda\}_\ast$ for some $\mathfrak{d}$ since after solving the equation for $W$ the following may happen
\begin{equation}\label{Path_to_Vas}
\Upsilon^{Vas}(\go,C)\neq \Upsilon^{Vas}(\go^\prime, C^\prime) \;\; \text{while}\;\; \{W,\Lambda\}_\ast(\go,C)=\{W,\Lambda\}_\ast(\go^\prime,C^\prime)\,.
\end{equation} 
for some $\go$, $\go^\prime$, $C$, $C^\prime$. Eq. \eqref{Path_to_Vas} implies that for the self-dual system $\omega$ and $C$ are related with $\go^\prime$ and $C^\prime$ by virtue of the (self-dual) gauge transformation while it is not true for Vasiliev system. This means that systems are given in different field frames (if two theories are equivalent in the first place). One can adjust the field frame by changing the $\mathfrak{d}$ in the system \eqref{dxWeq}-\eqref{class}. However as was already mentioned in section \ref{BurnMango} it is not clear which class of field redefinitions can be accomplished by consistently varying $\mathfrak{d}$.

To attack the problem of field redefinition consider a slightly modified version of \eqref{dxWeq1}-\eqref{LambdaDef} that captures all possible field redefinitions on the level of generating system. We change \eqref{Lambdaeq1} to
\begin{equation}\label{Lambdaeq2}
\dr_z \Lambda=\mathscr{F}(C,\ldots,C)(y|x)\ast \gamma\,,
\end{equation}
adjust the definition of $\Lambda$ \eqref{LambdaDef} as
\begin{equation}\label{LambdaDef2}
\Lambda:=\theta^\alpha z_\alpha\int_0^1 \dr \tau \, \tau \, e^{i\tau z_\alpha y^\alpha}\, \mathscr{F}(C,\ldots,C)(-\tau z|x)\,.
\end{equation}
and modify \eqref{Ceq1} as
\begin{equation}
\dr_x \mathscr{F}(C,\ldots, C)(y|x)\ast \gamma=\dr_z \{W,\Lambda\}_\ast\,.
\end{equation}
Here $\mathscr{F}(C,\ldots,C)$ is an arbitrary function of the original $C$-fields that corresponds to HS curvatures. Such a modified system will produce the following dynamics
\begin{equation}\label{dxWeq2}
\dr_x \go+\go\ast \go=\Upsilon(\go,\go,\mathscr{F}(C,\ldots,C))+\Upsilon(\go,\go,\mathscr{F}(C,\ldots,C)),\mathscr{F}(C,\ldots,C))+\ldots
\end{equation}
\begin{equation}\label{Ceq2}
\dr_x \mathscr{F}(C,\ldots,C)=\Upsilon(\go,\mathscr{F}(C,\ldots,C))+\Upsilon(\go,\mathscr{F}(C,\ldots,C),\mathscr{F}(C,\ldots C))+\ldots 
\end{equation}
Vertices $\Upsilon(\go,\go,\bullet,\ldots,\bullet)$, $\Upsilon(\go,\bullet,\ldots,\bullet)$ carry the same functional dependence as vertices that one obtains solving original system \eqref{dxWeq1}-\eqref{LambdaDef}. The latter implies that modification \eqref{Lambdaeq2}-\eqref{Ceq2} is consistent due to consistency of the original system and corresponds to the change of the field frame
\begin{equation}
C\rightarrow \mathscr{F}(C,\ldots ,C)\,.
\end{equation}
Change of the field frame for $\omega$ one can obtain using the freedom of the choice for solution to homogeneous equation solving \eqref{Weq11}.

The above discussion provides some indirect mechanisms that might relate self-dual theory and (anti)holomorphic truncation of 4d Vasiliev theory by virtue of the proper choice for $\mathfrak{d}$ or $\mathscr{F}(C,\ldots,C)$ but for the lowest order in nonlinearity, i.e. for vertices $\Upsilon(\go,C,C)$, the direct match is available. Moreover all of these vertices can be written down within the formalism of shifted homotopies. Vertices driven by $B_2^{\eta \, loc}$
\begin{multline}\label{B2loc}
B_2^{\eta \, loc}=\frac{\eta}{2}\int d^3 \tau_+\big[\delta^\prime(1-\sum_{i=1}^3 \tau_i)-iz_\alpha y^\alpha \delta(1-\sum_{i=1}^3\tau_i)\big]\exp\{i\tau_1 z_\alpha y^\alpha -i\tau_1 p_{1\alpha} p_2 {}^\alpha\}\times\\
\times C(-\tau_1 z+\tau_2 y)C(-\tau_1 z-\tau_3 y)k
\end{multline}
found in \cite{Vasiliev:2016xui}(see also \cite{Vasiliev:2017cae}) for generic background looks as follows
\begin{multline}\label{VgoCC}
\Upsilon^{\eta\, loc}_{\go CC}=-\frac{i\eta}{2}\int d^3 \tau_+\, \delta\big(1-\sum_{i=1}^3\tau_i\big)\, (y^\alpha t_\alpha)\, \exp\{-i(1-\tau_2)t_\alpha p_1 {}^\alpha+i\tau_2 t_\alpha p_2 {}^\alpha\}\times\\
\times \go\big((1-\tau_3)y\big)C(\tau_1 y)C\big((\tau_1-1)y\big)k\, ,
\end{multline}
\begin{multline}\label{VCCgo}
\Upsilon^{\eta\, loc}_{CC\go }=-\frac{i\eta}{2}\int d^3 \tau_+\, \delta\big(1-\sum_{i=1}^3 \tau_i\big)\, (y^\alpha t_\alpha)\, \exp\{-(1-\tau_2)p_{2\alpha}t^\alpha+i\tau_2 p_{1\alpha} t^\alpha\}\times\\
\times C\big((1-\tau_1)y\big)C(-\tau_1 y)\go\big((\tau_3-1)y\big)k\, ,
\end{multline}
\begin{multline}\label{VCgoC}
\Upsilon^{\eta}_{C\go C}=-\frac{i\eta}{2}\int d^3 \tau_+\, \delta\big(1-\sum_{i=1}^3 \tau_i\big)\, (y^\alpha t_\alpha)\, \exp\{-i\tau_2 p_{1\alpha} t^\alpha-i(1-\tau_2)t_\alpha p_2 {}^\alpha\}\times\\
\times C(\tau_3 y)\go(-\tau_1 y)C\big((\tau_3-1)y\big)k-\\
-\frac{i\eta}{2}\int d^3 \tau_+\, \delta \big(1-\sum_{i=1}^3 \tau_i\big)\, (y^\alpha t_\alpha)\exp\{-i(1-\tau_2)p_{1\alpha}t^\alpha-i\tau_2 t_\alpha p_2 {}^\alpha\}\times\\
\times C\big((1-\tau_3)y\big)\go(\tau_1 y)C(-\tau_3 y)k\,.
\end{multline}

Despite the fact that it is vely likely that one cannot obtain $B_2$ in the form \eqref{B2loc} using shifted homotopies, the vertices nonetheless can be written in the shifted homotopy formalism
\begin{equation}\label{v1}
\Upsilon^\eta_{\go CC}=\frac{\eta}{4i}\go \ast C \ast C\ast h_{y+p_2}\hmt_{p_1+2p_2}\hmt_{p_1+2p_2+t}\gamma\,,
\end{equation}
\begin{equation}
\Upsilon^\eta_{CC\go }=\frac{\eta}{4i}C\ast C\ast \go\ast h_{-y+p_1+2p_2+2t}\hmt_{p_2+t}\hmt_{p_2+2t}\gamma\, ,
\end{equation}
\begin{equation}\label{v3}
\Upsilon^\eta_{C\go C}=\frac{\eta}{4i}C\ast \go\ast C\ast \big(h_{-y+p_1+2t+2p_2}\hmt_{p_2+t}\hmt_{p_2}\gamma-h_{y+p_2}\hmt_{p_1+t+2p_2}\hmt_{p_1+2t+2p_2}\gamma\big)\,.
\end{equation}
Comparing these expressions with \eqref{first}-\eqref{last} one finds perfect match\footnote{Sign alterations in indices of homotopy operators and cohomology projectors emerge due to external Klein operator $k$ that is present in Vasiliev system which makes $\gamma$ the central element of the algebra with product \eqref{Vasiliev_pr}.}.
The $B_2$ field in the form \eqref{B2loc} was recently obtained within formalism of differential contracting homotopy \cite{Vasiliev:2023yzx}. The fact that vertices nonetheless can be written in terms of shifted homotopies implies that there should be some relations for specific choice of parameters for differential contracting homotopy.

\section{Conclusion}
The consistency of the generating system \cite{Didenko:2022qga}, which produces vertices in self-dual HS theory, is examined. We demonstrate the gap in the original proof of consistency that arises due to the absence of the Leibniz rule, thereby liberating all possible choices for the odd linear operator responsible for the evolution of master fields in $z^\alpha$. To address this issue, a twisted version of the original generating system \eqref{dxWeq}-\eqref{class} is considered. We demonstrate how it can lead to inconsistent vertices for the inappropriate choice of $\mathfrak{d}$. Nonetheless, the original system with the canonical $\dr_z=\theta^\alpha \frac{\partial}{\partial z^\alpha}$ indeed provides consistent vertices as shown by direct computation. 

The consistency analysis reveals previously unknown relations \eqref{star_new} between the limiting star product \eqref{star_infty} and the operators of shifted homotopy \eqref{h_def}, which have the form of the star-exchange identities originally discovered for a different product \eqref{Vasiliev_pr}. It shown that all interaction vertices resulting from system \cite{Didenko:2022qga} can be concisely expressed within the formalism of shifted homotopies \cite{Gelfond:2018vmi},\cite{Didenko:2018fgx}. In addition a generalization of the projection identity is found \eqref{proj_new}. This observation allows for the liberation of the $\Lambda$-field, rigidly defined in the original system by \eqref{LambdaDef}, by adding exact forms of the specific type \eqref{candide} that do not violate the projective properties of the resulting $\Lambda$. This opens the door for seeking a generalization of the system \eqref{dxWeq1}-\eqref{LambdaDef}, particularly those that include a mixed sector thus covering the full HS dynamics.

The connection with 4d Vasiliev theory is discussed, and a couple of scenarios are proposed that might relate the self-dual system to the (anti)holomorphic truncation of 4d Vasiliev theory on the level of generating systems. Since the system \eqref{dxWeq1}-\eqref{LambdaDef} with modifications \eqref{Lambdaeq2}-\eqref{Ceq2} captures all possible field redefinitions of the self-dual system, the search for a suitable $\mathscr{F}(C,\ldots,C)$ looks very promising. For example, one can explore developing some perturbation theory on $\mathscr{F}(C,\ldots,C)$ in powers of $C$ to relate $\Lambda$ with the $S$-field from the Vasiliev system. It is worth noting that for the lowest order in nonlinearity, namely for the vertices $\Upsilon(\go, C, C)$, there is a perfect match. Since all the vertices (modulo field redefinitions) of the system \cite{Didenko:2022qga} can be expressed in terms of the shifted homotopies, it is quite surprising that they align with those obtained from Vasiliev theory. Despite numerous attempts \cite{Didenko:2018fgx}, it seems unlikely to find a proper $B_2$ field within the framework of shifted homotopies. Nonetheless, the vertices driven by this particular form of the $B_2$ field \eqref{B2loc} can be formulated in terms of shifted homotopies and the corresponding cohomology projectors \eqref{v1}-\eqref{v3}. In a recent paper \cite{Vasiliev:2023yzx}, a different homotopy approach called the {\it differential contracting homotopy} was developed, which allows to obtain $B_2$ in the necessary form that leads to vertices with the minimal number of spacetime derivatives \eqref{VgoCC}-\eqref{VCgoC}. This observation suggests a connection between differential and shifted homotopies for specific choices of parameters.

It is very unlikely that one will be able to formulate any version of $\mathfrak{d}$ that differs from the canonical form with \eqref{C1_to_C2} imposed, which leads to consistent vertices for the system \eqref{dxWeq}-\eqref{class}. However, if one relaxes the condition \eqref{C1_to_C2}, then any element of $\mathbf{C}^1$ can serve as a proper candidate for $\Lambda$. This freedom may be utilized in $d$-dimensional system \cite{Didenko:2023vna} to subtract the necessary traces and put the system on-shell, analogously to the approach of \cite{Vasiliev:2025hfh}. Moreover, the freedom in $\mathfrak{d}$ might be used to build a Coxeter-like generalization (see \cite{Vasiliev:2018zer}) for the system proposed in \cite{Didenko:2022qga}.

\section*{Acknowledgement}
The author thanks Alexander Tarusov, Nikita Misuna, Dmitry Ponomarev, Slava Didenko and Mikhail Vasiliev for their helpful comments on the manuscript.  The author also gratefully acknowledges several years of conversations with Slava Didenko on various aspects of the problem discussed in the manuscript.

\newcounter{appendix}
\setcounter{appendix}{1}
\renewcommand{\theequation}{\Alph{appendix}.\arabic{equation}}
\addtocounter{section}{1} \setcounter{equation}{0}
 \renewcommand{\thesection}{\Alph{appendix}.}

\addcontentsline{toc}{section}{\,\,\,\,\, Appendix A}

\section*{Appendix A}\label{AppA}
Identity that allows to prove consistency for orderings $\go C \go C^N$ is the following
\begin{equation}\label{VIM}
h_{y+e}\big(\hmt_a \hmt_f \gamma \ast \pi(\Gamma(z,y))\big)=h_{-y+a}\hmt_{-y+f}\big(\Gamma(z,y) \ast \hmt_e \gamma\big)
\end{equation}
where $a$, $e$ and $f$ are $y$-independent and $\Gamma(z,y)$ is arbitrary function from class $\mathbf{C}^0$.  Generic representative of $\mathbf{C}^0$ class can be written in the form (cf \eqref{Classes})
\begin{equation}
\Gamma(z,y)=\int \mathscr{D}\rho \int_0^1 d\tau\, \frac{1-\tau}{\tau}\, e^{i\tau z_\alpha(y-B)^\alpha+i(1-\tau)y^\alpha A_\alpha -i\tau B_\alpha A^\alpha}\,.
\end{equation}
Here $A$ and $B$ might depend on various $\rho$s which are integrated over some compact area. Particular form of this dependence, integration measure and domain are not important so in what follows so integral over $\rho$s will be implicit and $\Gamma$ will be written simply as
\begin{equation}\label{C0_Gamma}
\Gamma(z,y)=\int_0^1 d\tau\, \frac{1-\tau}{\tau}\, e^{i\tau z_\alpha(y-B)^\alpha+i(1-\tau)y^\alpha A_\alpha -i\tau B_\alpha A^\alpha}\,.
\end{equation}
Divergence in denominator is fictional: one should treat the above expression as generating expression for class $\mathbf{C}^0$ and elements of $\mathbf{C}^0$ can be obtained by taking derivatives over $A$ and $B$. Derivative over $B$ brings $z_\alpha$ to pre-exponent and cancel the potentially dangerous pole.

For future computation explicit expression for $\hmt_a \hmt_f \gamma$ is needed
\begin{equation}
\hmt_a \hmt_f \gamma=-(z+a)_\alpha(z+f)^\alpha \int d^3\tau_+\, \delta(1-\tau_1-\tau_2-\tau_3) e^{i\tau_1 z_\alpha y^\alpha-i\tau_2 a_\alpha y^\alpha -i\tau_3 f_\alpha y^\alpha}\,.
\end{equation}
This particular expression can be found in \cite{Didenko:2018fgx} however it is not handy for limiting star-product computations, the better one is obtained after change of integration variables
\begin{equation}
\hmt_a \hmt_f \gamma=-(z+a)_\alpha (z+f)^\alpha \int_0^1 d\tau_1 \int_0^1 d\rho\, (1-\tau_1)\, e^{i\tau_1 z_\alpha y^\alpha-i(1-\tau_1)(\rho a+(1-\rho) f)_\alpha y^\alpha}\,.
\end{equation}
Automorphism acts on $\Gamma(z,y)$ as follows
\begin{equation}
\pi\big[\Gamma(z,y)\big]:=\Gamma(-z,-y)=\int_0^1 d\tau\, \frac{1-\tau}{\tau}\, e^{i\tau z_\alpha(y+B)^\alpha-i(1-\tau)y^\alpha A_\alpha -i\tau B_\alpha A^\alpha}\,,
\end{equation}
i.e. it changes $A$ to $-A$ and $B$ to $-B$.

Pretty much staightforward computation yields
\begin{multline}
\hmt_a \hmt_f \gamma \ast \pi\big[\Gamma(z,y)\big]=\\=-\int_0^1 d\mathcal{T}\int_0^1 d\sigma \int_0^1 d\rho\, \frac{1-\mathcal{T}}{1-\sigma}\, (z+A+f)_\alpha (f-a)^\alpha \exp\Big\{i\mathcal{T}z_\alpha \big(y+\sigma A+(1-\sigma) B-(1-\sigma)(\rho a+(1-\rho)f)\big)^\alpha-\\
-i(1-\mathcal{T})y^\alpha \big(\rho a (1-\rho)f+A\big)_\alpha-i(\rho a +(1-\rho)f)_\alpha A^\alpha+i\big(A-(1-\sigma)B\big)_\alpha \big(\rho a+ (1-\rho)f+A\big)^\alpha\Big\}
\end{multline}
Now we are in position to compute the l.h.s. of \eqref{VIM} which after some simple regrouping accounts the following form
\begin{multline}\label{VIM_l}
h_{y+e}\Big\{\hmt_a \hmt_f \gamma \ast \pi \big[\Gamma(z,y)\big]\Big\}=\\
=\int_0^1 d\widetilde{\mathcal{T}} \int_0^1 d\widetilde{\sigma} \int_0^1 d\rho \, \frac{1-\widetilde{\mathcal{T}}}{1-\widetilde{\sigma}}\, (y-A-a+e)^\alpha (f-a)^\alpha\, \exp\Big\{iy^\alpha\Big(-(1-\widetilde{\mathcal{T}}-\widetilde{\mathcal{T}}\widetilde{\sigma})A+\widetilde{\mathcal{T}}(1-\widetilde{\sigma})B-\\-(1-\widetilde{\mathcal{T}}\widetilde{\sigma})(\rho a +(1-\rho) f)\Big)_\alpha-i\widetilde{\mathcal{T}}y^\alpha e_\alpha+ie^\alpha\big(\widetilde{\mathcal{T}}\widetilde{\sigma} A+\widetilde{\mathcal{T}}(1-\widetilde{\sigma})B-\widetilde{\mathcal{T}}(1-\widetilde{\sigma})(\rho a +(1-\rho)f)\big)_\alpha+\\
+i(\rho a +(1-\rho) f)^\alpha \big((1-\widetilde{\mathcal{T}})A-\widetilde{\mathcal{T}}(1-\widetilde{\sigma})B\big)_\alpha-i\widetilde{\mathcal{T}}(1-\widetilde{\sigma})B_\alpha A^\alpha\Big\}\,.
\end{multline}

Now we are going to compute the r.h.s. of \eqref{VIM}. Straightforward calculation gives
\begin{multline}\label{G_times_g}
\Gamma(z,y)\ast \hmt_e \gamma=\\=\theta^\alpha \int_0^1 d\mathcal{T}\, \mathcal{T}\, (z-A+e)_\alpha \int_0^1 d\sigma \, \frac{1-\sigma}{\sigma}\, \exp\Big\{i\mathcal{T}z_\alpha\big(y-\sigma B+\sigma e+(1-\sigma)A\big)^\alpha+\\
+i(1-\mathcal{T})y^\alpha(A-e)_\alpha+iA_\alpha e^\alpha -i\mathcal{T}\big(\sigma B-A\big)_\alpha(A-e)^\alpha\Big\}\,.
\end{multline}
To proceed we need to apply shifted homotopy to previous expression recall it is defined ass follows
\begin{equation}
\hmt_{-y+f}\mathcal{F}(z,y|\theta):=(z-y+f)^\alpha \frac{\partial}{\partial \theta^\alpha}\int_0^1\frac{dt}{t}\, \mathcal{F}\big(tz-(1-t)(-y+f),y|t\theta\big)\,.
\end{equation}
Result of application is the following
\begin{multline}
\hmt_{-y+f}\Big(\Gamma(z,y)\ast \hmt_e \gamma\Big)=\\
=(z-y+f)^\alpha \int_0^1 dt \int_0^1 d\mathcal{T}\, \mathcal{T}\int_0^1 d\sigma \, \frac{1-\sigma}{\sigma}\, (z-A+e)_\alpha\, \exp\Big\{i\mathcal{T}\big(tz-(1-t)(-y+f)\big)_\alpha\big(y-\\
-\sigma B+\sigma e+(1-\sigma)A\big)^\alpha+i(1-\mathcal{T})y^\alpha(A-e)_\alpha+iA_\alpha e^\alpha-i\mathcal{T}(\sigma B-A)_\alpha (A-e)^\alpha\Big\}\,.
\end{multline}
Finally application of projector gives
\begin{multline}\label{VIM_r}
h_{-y+a}\hmt_{-y+f}\Big(\Gamma(z,y)\ast \hmt_e \gamma\Big)=\\
=\int_0^1 dt \int_0^1 d\mathcal{T}\, \mathcal{T}\int_0^1 d\sigma\, \frac{1-\sigma}{\sigma}\, (f-a)^\alpha(y-a-A+e)_\alpha\exp\Big\{iy^\alpha\Big(\mathcal{T}\sigma B-\big(1-\mathcal{T}(1-\sigma)\big)e+\\+\big(1-\mathcal{T}-\mathcal{T}(1-\sigma)\big)A-\mathcal{T}(t a+ (1-t)f)\Big)_\alpha+ie^\alpha\big((1-\mathcal{T})A+\mathcal{T}\sigma B-\sigma \mathcal{T}(ta+(1-t)f)\big)_\alpha+\\
+i\mathcal{T}(ta+(1-t)f)^\alpha (-\sigma B+(1-\sigma)A)_\alpha-i\mathcal{T}\sigma B_\alpha A^\alpha\Big\}\,.
\end{multline}
Expressions \eqref{VIM_l} and \eqref{VIM_r} are indeed coincide. To make this identity manifest one renames $t$ to $\rho$ in \eqref{VIM_r} and changes integration variables $\widetilde{\mathcal{T}}$ and $\widetilde{\sigma}$ in \eqref{VIM_l} according to
\begin{equation}
\widetilde{\mathcal{T}}=1-\mathcal{T}(1-\sigma)\,,\;\;\; \widetilde{\mathcal{T}}(1-\widetilde{\sigma})=\mathcal{T}\sigma\,.
\end{equation}

\renewcommand{\theequation}{\Alph{appendix}.\arabic{equation}}
\addtocounter{appendix}{1} \setcounter{equation}{0}
\addtocounter{section}{1}
\addcontentsline{toc}{section}{\,\,\,\,\, Appendix B}

\section*{Appendix B}\label{AppB}
Another important identity discovered during analysis of the consistency of the system \eqref{dxWeq1}-\eqref{LambdaDef} is
\begin{equation}\label{VIM_two}
\dr_z\big(\Gamma(z,y)\ast \hmt_e \gamma\big)=h_{y-e}\Gamma(z,y)\ast \gamma\;\;\; \forall \Gamma(z,y)\in \mathbf{C}^0\,
\end{equation}
which generalisises projection identity \eqref{Proj1},\eqref{Proj2}.
Indeed for $\Gamma(z,y)$ given by \eqref{C0_Gamma} one obtains \eqref{G_times_g} for product with $\hmt_e \gamma$ with $y$-independent $e$. Application of $\dr_z$ gives
\begin{multline}
\dr_z\big(\Gamma(z,y)\ast \hmt_e \gamma\big)=\\
=\frac{1}{2}\theta_\alpha\theta^\alpha \int_0^1 d\mathcal{T}\, \mathcal{T}\int_0^1 d\sigma\, \frac{1-\sigma}{\sigma}\Big[2+ (z-A+e)^\alpha (-i\mathcal{T})(y-\sigma B+\sigma e+(1-\sigma)A)_\alpha\Big] \times \\ \times\exp\Big\{i\mathcal{T}z_\alpha\big(y-\sigma B+\sigma e+(1-\sigma)A\big)^\alpha+
+i(1-\mathcal{T})y^\alpha(A-e)_\alpha+iA_\alpha e^\alpha -i\mathcal{T}\big(\sigma B-A\big)_\alpha(A-e)^\alpha\Big\}=\\
=\frac{1}{2}\theta_\alpha\theta^\alpha \int_0^1 d\sigma\, \frac{1-\sigma}{\sigma}\int_0^1 d\mathcal{T}\, \mathcal{T}(2+\mathcal{T}\frac{\partial}{\partial \mathcal{T}})\times\\
\times \exp\Big\{i\mathcal{T}z_\alpha\big(y-\sigma B+\sigma e+(1-\sigma)A\big)^\alpha+
+i(1-\mathcal{T})y^\alpha(A-e)_\alpha+iA_\alpha e^\alpha -i\mathcal{T}\big(\sigma B-A\big)_\alpha(A-e)^\alpha\Big\}=\\
=\frac{1}{2}\theta_\alpha \theta^\alpha \int_0^1 d\sigma\, \frac{1-\sigma}{\sigma}\exp\{iz_\alpha(y-\sigma B+\sigma e+(1-\sigma)A)^\alpha+iA_\alpha e^\alpha-i(\sigma B-A)_\alpha(A-e)^\alpha\}\,.
\end{multline}
Taking into account Klein-like properties of $\gamma$ \eqref{klein_like} and 
\begin{multline}
h_{y-e}\int_0^1 d\tau \, \frac{1-\tau}{\tau}\, \exp\{i\tau z_\alpha(y-B)^\alpha+i(1-\tau)y^\alpha A_\alpha-i\tau B_\alpha A^\alpha\}=\\
=\int_0^1 d\tau\, \frac{1-\tau}{\tau}\, \exp\{i\tau y_\alpha(B-e)^\alpha-i\tau e_\alpha B^\alpha-i\tau B_\alpha A^\alpha+i(1-\tau)y^\alpha A_\alpha\}\,
\end{multline}
one easily shows that \eqref{VIM_two} holds. Moreover in the same fashion one can prove analogous identity
\begin{equation}
\dr_z\big(\hmt_e\gamma\ast \Gamma(z,y)\big)=\gamma\ast h_{-y-e}\Gamma(z,y)\,
\end{equation}
for $y$-independent shift $e$.

To deduce projective property for $\dr_z(\Phi\ast\hmt_a\gamma\ast \Gamma)$ for generic $\Phi$ from $\mathbf{C}^0$
\begin{equation}\label{Phi_class}
\Phi(z,y)=\int_0^1 d\tau_1 \frac{1-\tau_1}{\tau_1} \exp\{i\tau_1 z_\alpha(y-C)^\alpha+i(1-\tau_1)y^\alpha D_\alpha-i\tau_1 D_\alpha C^\alpha\}\,
\end{equation}
we consider expression expression
\begin{equation}
\dr_z \dr_z\big(\hmt_b(\Phi\ast \hmt_a \gamma)\ast \Gamma\big)\equiv 0\,.
\end{equation}
Applying already known projective identities and resolution of identity it casts into
\begin{equation}
\dr_z\big(\Phi \ast \hmt_a \gamma \ast \Gamma\big)-h_{y-a}\Phi\ast h_{-y+b+\widehat{q}}\big(\pi[\Gamma]\big)\ast \gamma+\dr_z\big(\hmt_b(\Phi\ast \hmt_a \gamma)\ast \dr_z\Gamma\big)\equiv 0\,.
\end{equation}
Here $\widehat{q}$ is the derivative over full holomorphic argument of $h_{y-a}\Phi$.
\begin{equation}
\dr_z\big(\Phi \ast \hmt_a \gamma \ast \Gamma\big)=h_{y-a}\Phi\ast h_{-y+b+\widehat{q}}\big(\pi[\Gamma]\big)\ast \gamma-\dr_z\big(\hmt_b(\Phi\ast \hmt_a \gamma)\ast \dr_z\Gamma\big)\,.
\end{equation}
Note the l.h.s. does not depend on $b$ so it can be chosen arbitrary on the r.h.s. It may even depend on $C$ and $D$ in \eqref{Phi_class}. Using this freedom we can make last term vanish (see \eqref{G_times_g}). Indeed
\begin{multline}\label{class_delta2}
\hmt_b\big(\Phi\ast \hmt_a\gamma\big)=-(z+b)^\alpha(D-a+b)_\alpha\int_0^1 d\tau\, (1-\tau)\int_0^1 d\rho\int_0^1 d\sigma \frac{1-\sigma}{\sigma}\times\\
\times \exp\{i\tau z_\alpha\big(y-\sigma C+\sigma b+(1-\sigma)D\big)^\alpha+i(1-\tau)y^\alpha\big((1-\rho)(D-a)-\rho b\big)_\alpha-\\
-i\tau\big(\sigma C-\sigma a-(1-\sigma)D\big)_\alpha\big((1-\rho)(C-a)-\rho b\big)^\alpha-\\
-i\rho \big(\sigma C-\sigma a-(1-\sigma)D\big)_\alpha (D-a+b)^\alpha+iD_\alpha a^\alpha\}\,.
\end{multline}
The latter implies that
\begin{equation}
\dr_z\big(\hmt_b(\Phi\ast \hmt_a \gamma)\ast \dr_z \Gamma\big)=0\;\;\;\; for\;\;\;\; b=a-D\,.
\end{equation}
So we are left with
\begin{equation}\label{proj_new3}
\dr_z \big(\Phi\ast \hmt_a \gamma \ast \Gamma\big)=h_{y-a}\Phi \ast h_{-y+a-D+\widehat{q}}\big(\pi[\Gamma]\big)\ast \gamma\,.
\end{equation}

\renewcommand{\theequation}{\Alph{appendix}.\arabic{equation}}
\addtocounter{appendix}{1} \setcounter{equation}{0}
\addtocounter{section}{1}
\addcontentsline{toc}{section}{\,\,\,\,\, Appendix C}

\section*{Appendix C}\label{AppC}

To demonstrate inconsistency of the system \eqref{dxWeq}-\eqref{class} with $\mathfrak{d}$ given by \eqref{definition} we examine order $\go\go CC$ in zero-form sector. We use \eqref{modified} expression for the $\Lambda$ field assuming that $\varepsilon$ is linear in $C$-field. To proceed we solve equations of generating system. We start with \eqref{Weq}, for $W^\prime_{\go C}$ we need to solve
\begin{equation}
\dr_z W^\prime_{\go C}=-\go\ast \Lambda-\dr_x\Lambda-\go\ast \dr_z \varepsilon-\dr_x\dr_z \varepsilon\,.
\end{equation}
Here $\Lambda$ is given by \eqref{LambdaDef}. It is easy to rewrite last two terms as total derivative
\begin{equation}
\dr_z W^\prime_{\go C}=-\go\ast\Lambda-\dr_x\Lambda+\dr_z(\go\ast \varepsilon+\dr_x \varepsilon)\,.
\end{equation}
Particular solution has the form
\begin{equation}
W^\prime_{\go C}=\go \ast C\ast \hmt_{p+t}\hmt_p\gamma+\go\ast \varepsilon+\dr_x \varepsilon\,.
\end{equation}
Even though expression for $W^\prime_{\go C}$ differs from \eqref{WwC} resulting vertices are the same. Indeed, additional contribution to vertex that comes from $\go\ast \varepsilon+\dr_x\varepsilon$ for ordering $\go\go C$ looks as follows
\begin{equation}
-\dr_x\big(\go\ast \varepsilon+\dr_x \varepsilon\big|_{\go C}\big)-\go\ast\big(\go\ast\varepsilon+\dr_x \varepsilon\big|_{\go C}\big)=0\,.
\end{equation}

In contrast to one-forms vertices for zero-form sector are different
\begin{equation}
\Upsilon^\prime_{\go C C}\ast \gamma=\mathfrak{d}\big[(W_{\go C}+\go\ast \varepsilon+\dr_x \varepsilon)\ast (\Lambda+\dr_z \varepsilon)\big]\,.
\end{equation}
There is a vertex that comes from original system namely
\begin{equation}
\Upsilon_{\go CC}\ast \gamma=\dr_z (W_{\go C}\ast \Lambda)\,.
\end{equation}
Here we used the fact that product $W_{\go C}\ast \Lambda$ is projectable and so we can change $\mathfrak{d}$ to $\dr_z$. Additional contribution to the vertex is given by
\begin{equation}
\widetilde{\Upsilon}_{\go CC}\ast \gamma=\dr_z (W_{\go C}\ast \dr_z \varepsilon)+\dr_z\big[(\go\ast \varepsilon+\dr_x \varepsilon)\ast \Lambda\big]+\mathfrak{d}\big[(\go\ast \varepsilon+\dr_x \varepsilon)\ast \dr_z \varepsilon\big]\,.
\end{equation}
Using projection identities \eqref{proj_new},\eqref{proj_new2} we can show that additional contribution acquires the following compact form
\begin{multline}
\widetilde{\Upsilon}_{\go CC}\ast \gamma=\Big[-\go\ast h_{y+p_2}(\varepsilon)\ast C-h_{y+p_2}\big(\dr_x\varepsilon\big|_{\go C}\big)\ast C+\go\ast C\ast h_{-y+p_1}\big(\pi[\varepsilon]\big)-\\
-\go \ast C\ast h_{-y+p_1+t_1}\big(\pi[\varepsilon]\big)\Big]\ast \gamma
+\mathfrak{d}\big[(\go\ast \varepsilon+\dr_x \varepsilon)\ast \dr_z \varepsilon\big]\,.
\end{multline}
To check if consistency holds for ordering $\go\go CC$ we need to inspect the following expression
\begin{equation}
\Big[\underbrace{-\Upsilon_{\go CC}\ast C+\go\ast \Upsilon_{\go CC}+\dr_x \Upsilon_{\go CC}\Big|_{\go\go CC}}+\dr_x\widetilde{\Upsilon}_{\go C C}\Big|_{\go\go CC}+\go\ast \widetilde{\Upsilon}_{\go CC}\Big]\ast \gamma\,.
\end{equation}
Here underbraced terms vanish due to consistency of original system \eqref{dxWeq1}-\eqref{LambdaDef} proven in section \ref{kinematic}. For the remaining terms one can easily show that it casts to (other contributions vanish)
\begin{equation}\label{inconsistency}
\mathfrak{d}\Big[(\go\ast \go\ast \varepsilon+\go\ast \dr_x \varepsilon)\ast \dr_z \varepsilon\Big]+\go\ast \mathfrak{d}\Big[(\go\ast \varepsilon+\dr_x \varepsilon)\ast \dr_z \varepsilon\Big]
\end{equation}
For generic $\varepsilon\in \mathbf{C}^0$ non of these terms are projectable and hence nonzero shift $a$ for $y$ argument in definition of $\mathfrak{d}$ (see \eqref{definition}) affects the form of resulting expression. Schematically \eqref{inconsistency} can be written as
\begin{equation}
\big[\go(\ldots +a)X(y,a)-\go(\ldots)X(y,a)\big]\ast \gamma\neq 0\,
\end{equation}
where the periods in the arguments of $\go$ written explicitly symbolize the same expression in both terms. This results in an inconsistency for the ordering $\go\go CC$ for the system \eqref{dxWeq}-\eqref{class} with $\mathfrak{d}$ defined as in \eqref{definition} and the $\Lambda$-field shifted by the exact form \eqref{modified}. Although setting $a$ to zero naively restores consistency, it is only restored for this specific order of perturbation theory. For $a=0$, the structures $\go \go C\ldots C$ exhibit inconsistency starting from the third order in $C$, but the computations become significantly more involved.

\end{document}